\documentclass[9pt,nocopyrightspace,preprint]{sigplanconf}

\usepackage[utf8]{inputenc}
\usepackage[T1]{fontenc}
\usepackage{amsmath}
\usepackage{amssymb}
\usepackage[usenames]{color}
\usepackage{graphicx}
\usepackage{xcolor}
\usepackage{listings}
\usepackage[scaled=0.8]{beramono}
\usepackage{multicol}
\usepackage{url}
\usepackage{breakurl}
\usepackage{hyperref}
\usepackage[compact]{titlesec}
\usepackage{framed}
\usepackage{mdframed}
\usepackage{flushend}

\usepackage{math}
\usepackage{fleqn}
\usepackage{latexsym}
\usepackage{bcprules}
\usepackage{prooftree}

\lstdefinelanguage{Scala}%
{morekeywords={abstract,%
  case,catch,char,class,%
  def,else,extends,final,finally,for,%
  if,import,implicit,%
  match,
  new,null,%
  object,override,%
  package,private,protected,public,%
  for,public,return,super,sealed,%
  throw,trait,try,type,%
  val,var,%
  with,while,%
  yield%
  },%
  sensitive,%
  morecomment=[l]//,%
  morecomment=[s]{/*}{*/},%
  morestring=[b]",%
  morestring=[b]',%
  showstringspaces=false%
}[keywords,comments,strings]%

\definecolor{listingbg}{RGB}{240, 240, 240}

\newcommand{\keywordstyle}[1]{\bfseries{#1}}

\lstnewenvironment{listing}{\lstset{language=Scala,mathescape=true}}{}
\lstnewenvironment{listingtiny}{\lstset{language=Scala,mathescape=truebasicstyle=\scriptsize\ttfamily}}{}

\newcommand{\comment}[1]{}

\lstdefinelanguage{DOT}%
{morekeywords={val,new},%
  sensitive,%
  morecomment=[l]//,%
  morecomment=[s]{/*}{*/},%
  morestring=[b]",%
  morestring=[b]',%
  showstringspaces=false%
}[keywords,comments,strings]%

\newcommand{\figurebox}[1]
        {\fbox{\begin{minipage}{\textwidth} #1 \medskip\end{minipage}}}

\newcommand{\boxfig}[3]
        {\begin{figure*}\figurebox{#3\caption{\label{#1}#2}}\end{figure*}}

\newlength{\trulemargin}
\newlength{\trulewidth}
\newlength{\srulewidth}
\newenvironment{trules}{$\vspace{0.5em}\ba{p{\trulemargin}@{~}p{\trulewidth}@{~}p{\trulemargin}}}{\ea$}
\newenvironment{srules}{$\vspace{0.5em}\ba{p{\trulemargin}@{~}p{\srulewidth}}}{\ea$}

\newcommand{\gap}{\quad\quad}

\newcommand{\ba}{\begin{array}}
\newcommand{\ea}{\end{array}}

\newcommand{\ei}{\end{array}}
\newcommand{\bcases}{\left\{\begin{array}{ll}}
\newcommand{\ecases}{\end{array}\right.}

\newcommand{\ts}{\,\vdash\,}
\newcommand{\tS}{\,\vDash\,}

\newcommand{\spcomma}{~,~}

\newcommand{\sub}{<:}

\newcommand{\reduces}{\;\rightarrow\;}

\newcommand{\tand}{\wedge}
\newcommand{\tor}{\vee}

\newcommand{\judgement}[2]{{\bf #1} \hfill #2}

\newcommand{\seq}[1]{\colorlet{default}{.}\color{gray}\overline{\color{default}#1}\color{default}}

\newcommand{\Top}{\top}
\newcommand{\Bot}{\bot}

\newcommand{\clos}[1]{\left<#1\right>}

\graphicspath{{figures/}}

\begin{document}


\title{From F to DOT: Type Soundness Proofs with Definitional Interpreters}

\titlebanner{\\[-2ex] --- Technical Report --- \\
\scriptsize{\texttt {Version 2.0: revised and extended January 2016. Initially published July 2015.}}}

\authorinfo{
Tiark Rompf$^{\;\ast}$
\and Nada Amin$^\dagger$
} {
$^\ast$Purdue University: \texttt{\{first\}@purdue.edu}\qquad \\
$^\dagger$EPFL: \texttt{\{first.last\}@epfl.ch}
\vspace{-13mm}
}


\lstMakeShortInline[keywordstyle=,%
              flexiblecolumns=false,%
              mathescape=false,%
              basicstyle=\tt]@

\maketitle

\begin{abstract}

Scala's type system unifies aspects of ML modules, object-oriented, and 
functional programming. The Dependent Object Types (DOT) family of calculi 
has been proposed as a new theoretic foundation for Scala and similar 
expressive languages. 
Unfortunately, it is not clear how DOT relates to well-studied type systems 
from the literature, and type soundness has only been established for very
restricted subsets.
%
In fact, it has been shown that important Scala features such as type refinement 
or a subtyping relation with lattice structure break at least one key metatheoretic 
property such as environment narrowing or subtyping transitivity, which are usually 
required for a type soundness proof.

The first main contribution of this paper is to demonstrate how, perhaps surprisingly, 
even though these properties are lost in their full generality, a rich DOT calculus that 
includes both type refinement and a subtyping lattice with intersection types can still be 
proved sound. The key insight is that narrowing and subtyping transitivity only need to 
hold for \emph{runtime objects}, but not for code that is never executed. Alas, the 
dominant method of proving type soundness, Wright and Felleisen's syntactic approach, is 
based on term rewriting, which does not a priori make a distinction between runtime and 
type assignment time.

The second main contribution of this paper is to demonstrate how type soundness proofs for 
advanced, polymorphic, type systems can be carried out with an operational semantics based 
on high-level, definitional interpreters, implemented in Coq. We present the first 
mechanized soundness proof in this style for System F$_{\sub}$ and several extensions, 
including mutable references. Our proofs use only straightforward induction, which is
another surprising result, as the combination of big-step semantics, mutable references, 
and polymorphism is commonly believed to require co-inductive proof techniques.

The third main contribution of this paper is to show how DOT-like calculi emerge from 
straightforward generalizations of the operational aspects of F$_{\sub}$, exposing a
rich design space of calculi with path-dependent types inbetween System F and DOT,
which we collectively call System D. Armed with the insights gained from the 
definitional interpreter semantics, we also show how equivalent small-step 
semantics and soundness proofs in the style of Wright and Felleisen can be 
derived for these systems.




\end{abstract}

\section{Introduction}\label{sec:intro}

Modern expressive programming languages such as Scala integrate and generalize
concepts from functional programming, object oriented programming and ML-style 
module systems \cite{DBLP:journals/cacm/OderskyR14}. While most of these features 
are understood on a theoretical level in isolation, their combination is not, 
and the gap between formal type theoretic models and what is implemented in 
realistic languages is large.

In the case of Scala, developing a sound formal model that captures
a relevant subset of its type system has been an elusive quest for more than a 
decade. After many false starts, the first mechanized
soundness proof for a calculus of the DOT (Dependent Object Types) \cite{dotfool}
family was finally presented in 2014 \cite{DBLP:conf/oopsla/AminRO14}.

The calculus proved sound by Amin et al.\ \cite{DBLP:conf/oopsla/AminRO14} 
is $\mu$DOT, a core calculus that distills the essence of Scala's objects 
that may contain \emph{type members} in addition to methods and fields, 
along with \emph{path-dependent types}, which are used to access such type 
members. 
$\mu$DOT models just these two features--records 
with type members and type selections on variables--and nothing else.
This simple system already captures some essential programming patterns,
and it has a clean and intuitive theory. In particular, it satisfies the 
intuitive and mutually dependent properties of environment narrowing and 
subtyping transitivity, which are usually key requirements for a soundness 
proof. 

Alas, Amin et al.\ also showed that adding other important Scala features 
such as type refinement, mixin composition, or just a bottom type
breaks at least one of these properties, which 
makes a soundness proof much harder to come by. 

The first main contribution of this paper is to demonstrate how, 
perhaps surprisingly, even though these properties are lost in their
full generality, a richer DOT calculus that includes both type refinement
and a subtyping lattice with full intersection and union types can still 
be proved sound.
The key insight is that narrowing and subtyping transitivity only 
need to hold for types assigned to \emph{runtime objects}, but not for 
arbitrary code that is never executed. 

But how can we leverage this insight in a type safety proof? The dominant
method of proving type soundness, Wright and Felleisen's syntactic 
approach~\cite{DBLP:journals/iandc/WrightF94}, relies on establishing
a type preservation (or subject reduction) property of a small-step 
operational semantics based on term rewriting, which does not distinguish 
terms given by the programmer from (partially) reduced terms resulting
from prior reductions.
The cornerstone of most soundness proofs is a substitution lemma
that establishes type preservation of function application via 
$\beta$-reduction. But in the case of DOT, this very property does not hold!

Among the many desirable properties of the syntactic approach to type
soundness is its
generality: almost anything can be encoded as a term rewriting
system, and dealt with in the same uniform way using ordinary
inductive techniques, as opposed to requiring different and complicated
proof techniques for different aspects of the semantics (state,
control, concurrency, ...)~\cite{DBLP:journals/iandc/WrightF94}. 
But the downside is that few realistic languages \emph{actually} 
preserve types across arbitrary substitutions,
and that \emph{no} realistic language implementation actually proceeds
by term rewriting. Thus, coaxing the syntax, reduction relation, and 
type assigment rules to enable subject reduction requires ingenuity
and poses the question of adequacy: are we really formalizing the
language we think we do?

In this paper, we present a different approach. Our second main contribution 
is to demonstrate how type soundness for advanced, polymorphic, 
type systems can be proved with respect to an operational semantics based on 
high-level, definitional interpreters, implemented directly in a total 
functional language like Coq. While this has been done before for
very simple, monomorphic, type systems~\cite{siek3lemmas,DBLP:conf/icfp/Danielsson12}, 
we are the first to demonstrate that, with some additional machinery,  
this approach scales to realistic, polymorphic type 
systems that include subtyping, abstract types, and types with binders.

Our proofs use only straightforward induction, even if we add features
like mutable references. This in itself is a surprising result,
as the combination of big-step semantics, mutable references, 
and polymorphism is commonly believed to require co-inductive proof techniques.

We develop our mechanization of DOT gradually: we first present a 
mechanized soundness proof for System F$_{\sub}$ based on a definitional 
interpreter that includes \emph{type values} as runtime objects. 
Perhaps surprisingly, many of the DOT challenges already
arise in this simpler setting, in particular because F$_{\sub}$
already requires relating abstract types across contexts at runtime.

From there, as our third main contribution, we illustrate how DOT-like 
calculi emerge as generalizations of the static typing rules to fit 
the operational typing aspects of the F$_{\sub}$ based system---in 
some cases almost like removing artifical restrictions. 
By this, we put DOT on a firm theoretical foundation based on
existing, well-studied, type systems, and we expose a rich design 
space of calculi with path-dependent types 
inbetween System F and DOT, which we collectively call System D. 

Based on the development of the definitional interpreter semantics, 
we also show how equivalent rewriting semantics and soundness 
proofs in the style of Wright and Felleisen can be 
derived for these systems.

We believe that this angle, taking the runtime aspects of (abstract and
even polymorphic) types as a starting point of investigation,
is an interesting but not very well developed approach that nicely
complements the more traditional `start from static terms' approach
and may lead to interesting novel type system developments.
A key take-away is that the static aspects of a 
type system most often serve to \emph{erect} abstraction boundaries 
(e.g.\ to enforce representation independence), whereas the
dynamic aspects of a type system must serve to relate types 
\emph{across} such abstraction boundaries.

The paper is structured around the three main contributions:
\begin{itemize}

\item We present the first sound calculus of the 
DOT family that includes type refinement and a subtyping
lattice with intersection and union types:
first informally with examples (Section~\ref{sec:main1}),
then formally (Section~\ref{sec:main2}), and
we discuss the key properties that make a soundness
proof difficult (Section~\ref{sec:main3}).

\item We discuss liminations of the syntactic approach to
soundness and demonstrate that a proof strategy based on
high-level definitional interpreters (Section~\ref{sec:defint})
scales to advanced polymorphic type systems, presenting 
the first soundness proof for F$_{\sub}$ in this style 
(Section~\ref{sec:fsub}). 

\item We demonstrate how DOT emerges through extensions 
of F$_{\sub}$, and we present new foundations for DOT
through newly discovered intermediate systems such as D$_{\sub}$, 
which may also be of independent interest (Section~\ref{sec:dsub}). 
Finally, we derive equivalent reduction semantics and soundness proofs 
from the definitional interpreter construction 
(Section~\ref{sec:smallstep}).

\end{itemize}

We discuss related work in Section~\ref{sec:related} and offer 
concluding remarks in Section~\ref{sec:conclusion}.
Our mechanizations of F$_{<:}$, D$_{<:}$, DOT, and their
variations are available from:

\vspace{2pt}
{\hspace{1cm}{\small\url{http://github.com/tiarkrompf/minidot}}}


\section{Types in Scala and DOT}\label{sec:main1}


Scala is a large language that combines features from
functional programming, object-oriented programming and
module systems. Scala unifies many of these features
(e.g.\ objects, functions, and modules) 
\cite{DBLP:journals/cacm/OderskyR14}
but still contains a large set of distinct kinds of 
types. These can be broadly classified \cite{troubleWithTypes} into
\emph{modular} types:
\begin{lstlisting}[keywords={}]
                     named type: scala.collection.BitSet
                  compound type: Channel with Logged
                   refined type: Channel { def close(): Unit }
\end{lstlisting}
And \emph{functional} types :
\begin{lstlisting}[keywords={}]
             parameterized type: List[String]
               existential type: List[T] forSome { type T }
             higher-kinded type: List
\end{lstlisting}

While this variety of types enables programming 
styles appealing to programmers with different backgrounds
(Java, ML, Haskell, ...), not all of them are essential.
Further unification and an economy of concepts can be achieved by 
reducing functional to modular types as follows:
\begin{lstlisting}[keywords={}]
           class List[Elem] {} $\rightarrow$ class List { type Elem }
                  List[String] $\rightarrow$ List { type Elem = String }
    List[T] forSome { type T } $\rightarrow$ List
                          List $\rightarrow$ List
\end{lstlisting}

This unification is the main thrust of the calculi of the DOT family.
A further thrust is to replace Scala's compound types @A with B@ with
proper intersection types @A & B@.
Before presenting our DOT variant in a formal setting in Section~\ref{sec:main2},
we introduce the main ideas with some high-level programming examples.

\paragraph{Objects and First Class Modules}

In Scala and in DOT, every piece of data is conceptually an object and every 
operation a method call. This is in contrast to functional languages in the ML 
family that are stratified into a core language and a module system. 
Below is an implementation of a functional list data structure:
\begin{lstlisting}
val listModule = new { m =>
  type List = { this =>
    type Elem
    def head(): this.Elem
    def tail(): m.List & { type Elem <: this.Elem }
  }
  def nil() = new { this =>
    type Elem = Bot
    def head() = error()
    def tail() = error()
  }
  def cons[T](hd: T)(tl: m.List & { type Elem <: T }) = new { this =>
    type Elem = T
    def head() = hd
    def tail() = tl
  }
}
\end{lstlisting}
The actual @List@ type is defined inside a container @listModule@,
which we can think of as a first-class module. In an extended DOT calculus
@error@ may signify an `acceptable' runtime error or exception
that aborts the current execution and transfers control to an
exception handler. In the case that we study, without such 
facilities, we can model the abortive behavior of @error@ as a 
non terminating function, for example @def error(): Bot = error()@.

\paragraph{Nominality through Ascription}

In most other settings (e.g. object-oriented subclassing, ML module 
sealing), nominality is enforced when objects are declared or 
constructed. Here we can just assign a more abstract
type to our list module:
\begin{lstlisting}
type ListAPI = { m =>
  type List <: { this =>
    type Elem
    def head(): this.Elem
    def tail(): m.List & { type Elem <: this.Elem }
  }
  def nil(): List & { type Elem = Bot }
  def cons[T]: T => 
    m.List & { type Elem <: T } => 
      m.List & { type Elem <: T }
}
\end{lstlisting}
Types @List@ and @Elem@ are abstract, and thus exhibit nominal
as opposed to structural behavior.
Since modules are just objects, it is perfectly possible to pick
different implementations of an abstract type based on runtime 
parameters.
\begin{lstlisting}
val mapImpl = if (size < 100) ListMap else HashMap
\end{lstlisting}

\paragraph{Polymorphic Methods}

In the code above, we have still used the functional notation
@cons[T](...)@ for parametric methods. We can desugar the
type parameter $T$ into a proper method parameter $t$ with 
a modular type, and at the same time desugar the multi-argument 
function into nested anonymous functions:
\begin{lstlisting}
def cons(t: { type T }) = ((hd: t.T) => ... )
\end{lstlisting}
References to $T$ are replaced by a path dependent type $t.T$. 
We can further desugar the anonymous functions into objects
with a single @apply@ method:
\begin{lstlisting}
def cons(t: { type T }) = new {
  def apply(hd: t.T) = new {
    def apply(tl: m.List & { type Elem <: t.T }) = new { this => 
      type Elem = t.T
      def head() = hd
      def tail() = tl
    }}}
\end{lstlisting}

\paragraph{Path-Dependent Types}
Let us consider another example to illustrate path-dependent types: 
a system of services, each with a specific type of 
configuration object. Here is the abstract interface:
\begin{lstlisting}
type Service {
  type Config
  def default: Config
  def init(data: Config):Unit
}
\end{lstlisting}
We now create a system consisting of a database and an
authentication service, each with their respective configuration
types:
\begin{lstlisting}
type DBConfig { def getDB: String }
type AuthConfig { def getAuth: String }
val dbs = new Service { 
  type Config = DBConfig
  def default = new DBConfig { ... }
  def init(d:Config) = d.getDB 
}
val auths = new Service { 
  type Config = AuthConfig
  def default = new AuthConfig { ... }
  def init(d:Config) = d.getAuth 
}
\end{lstlisting}
We can inititialize @dbs@ with a new @DBConfig@, and @auths@ with 
a new @AuthConfig@, but not vice versa. This is because each object has
its own specific @Config@ type member and thus, @dbs.Config@ and 
@auths.Config@ are distinct \emph{path dependent} types.
Likewise, if we have a service @lambda: Service@ without further
refinement of its @Config@ member, we can still call
@lam.init(lam.default)@
but we cannot create a @lam.Config@ value directly, because @Config@
is an abstract type in @Service@.

\paragraph{Intersection and Union Types}
At the end of the day, we want only one centralized configuration for 
our system, and we can create one by assigning an intersection type:
\begin{lstlisting}
val globalConf: DBConfig & AuthConfig = new {
  def getDB = "myDB"
  def getAuth = "myAuth"
}
\end{lstlisting}
Since @globalConf@ corresponds to both @DBConfig@ and @AuthConfig@, we can use
it to initialize both services:
\begin{lstlisting}
dbs.init(globalConf)
auths.init(globalConf)
\end{lstlisting}
But we would like to abstract even more.

With the @List@ definition presented earlier, we can build a list of 
services (using @::@ as syntactic sugar for @cons@):
\begin{lstlisting}
val services = auths::dbs::Nil
\end{lstlisting}
We define an initialization function for a whole list of services:
\begin{lstlisting}
def initAll[T](xs:List[Service { type Config >: T }])(c: T) = 
  xs.foreach(_ init c)
\end{lstlisting}
Which we can then use as:
\begin{lstlisting}
initAll(services)(globalConf)
\end{lstlisting}

How do the types play out here? The definition of @List@ and 
@cons@ makes the type member @Elem@ covariant. Thus, the type 
of @auths::dbs::Nil@ corresponds to 
\begin{lstlisting}
List & {
  type Elem = Service & { 
    type Config: (DBConfig & AuthConfig) .. (DBConfig | AuthConfig) 
  }
}
\end{lstlisting}
This means that we can treat the @Config@ member as lower bounded by
@DBConfig & AuthConfig@, so passing an object of that type to @init@ is
legal.

\paragraph{Records and Refinements as Intersections}

Subtyping allows us to treat a type as a less precise one.
Scala provides a dual mechanism that enables us to create a
more precise type by \emph{refining} an existing one.
\begin{lstlisting}
type PersistentService = Service { a =>
  def persist(config: a.Config)
}
\end{lstlisting}

To express the type @PersistentService@ by desugaring the refinement into an
intersection type, we need a ``self'' variable (here @a@) to close
over the intersection type, in order to refer to the abstract type
member @Config@ of @Service@:
\begin{lstlisting}
type PersistentService = { a => Service & {
  def persist(config: a.Config)
}}
\end{lstlisting}

Our variant of DOT uses intersections also to model records with multipe
type, value, or method members:
\begin{lstlisting}
{ def foo(x)=..; def bar(y)=..}  =  { def foo(x)=..} & { def bar(y)=..}
\end{lstlisting}
With this encoding of records, we benefit again from an economy of concepts.

\section{Formal Model of DOT}\label{sec:main2}

\boxfig{fig:dot1}{DOT: Syntax and Type System}{
\judgement{Syntax}{}
\begin{center}
$
\ba{l|l|l}
\ba{ll}
x,y,z         & \mbox{Variable}\gap\gap\gap~\\
L             & \mbox{Type label}\\
l             & \mbox{Value label}\\
m             & \mbox{Method label}\\[1cm]
\ea
&
\ba{ll}
S,T,U ::=     & \mbox{Type}\\
\gap \top     & \gap\mbox{top type}\\
\gap \bot     & \gap\mbox{bottom type}\\
\gap L:S..U   & \gap\mbox{type member}\\
\gap l:T      & \gap\mbox{value member}\\
\gap m(x:S):U^x & \gap\mbox{method member}\\
\gap x.L      & \gap\mbox{type selection}\\
\gap \{ z \Rightarrow T^z \} & \gap\mbox{recursive self type}\\
\gap T \tand T & \gap\mbox{intersection type}\\
\gap T \tor T & \gap\mbox{union type}
\ea
&
\ba{ll}
t ::=         & \mbox{Term}\\
\gap x        & \gap\mbox{variable reference}\\
\gap \{ z \Rightarrow \seq{d} \} & \gap\mbox{new instance}\\
\gap {t.l}    & \gap\mbox{field selection}\\
\gap {t.m(t)} & \gap\mbox{method invocation}\\
d ::=         & \mbox{Initialization}\\
\gap L = T    & \gap\mbox{type initialization}\\
\gap l = t    & \gap\mbox{field initialization}\\
\gap m(x) = t & \gap\mbox{method initialization}
\ea
\ea
$
\end{center}
\hrule\vspace{1ex}

\setlength{\columnseprule}{0.4pt}
\setlength{\columnsep}{8pt}
\begin{multicols}{2}

\judgement{Subtyping}{\fbox{$\Gamma \ts S \sub U$}}

\typicallabel{}
\smallrulenames
\setlength{\columnseprule}{0pt}
\begin{multicols}{2}[Lattice structure]

\infax[Bot]{\Gamma \ts \bot \sub T}

\infrule[And11]
{\Gamma \ts T_1 \sub T}
{\Gamma \ts T_1 \tand T_2 \sub T}

\infrule[And12]
{\Gamma \ts T_2 \sub T}
{\Gamma \ts T_1 \tand T_2 \sub T}

\infrule[And2]
{\Gamma \ts T \sub T_1 \spcomma T \sub T_2}
{\Gamma \ts T \sub T_1 \tand T_2}

\infax[Top]{\Gamma \ts T \sub \top}

\infrule[Or21]
{\Gamma \ts T \sub T_1}
{\Gamma \ts T \sub T_1 \tor T_2}

\infrule[Or22]
{\Gamma \ts T \sub T_2}
{\Gamma \ts T \sub T_1 \tor T_2}

\infrule[Or1]
{\Gamma \ts T_1 \sub T \spcomma T_2 \sub T}
{\Gamma \ts T_1 \tor T_2 \sub T}
\end{multicols}

\bigskip
Type, field and method members

\infrule[Fld]
{\Gamma \ts T_1 \sub T_2}
{\Gamma \ts l:T_1 \sub l:T_2}

\infrule[Mem]
{\Gamma \ts S_2 \sub S_1 \spcomma U_1 \sub U_2}
{\Gamma \ts L:S_1..U_1 \sub L:S_2..U_2}

\infrule[Fun]
{\Gamma \ts S_2 \sub S_1\\ \Gamma,x:S_2 \ts U_1^x \sub U_2^x}
{\Gamma \ts m(x:S_1):U_1^x \sub m(x:S_2):U_2^x}

\bigskip
Path selections

\infax[SelX]{\Gamma \ts x.L \sub x.L}

\begin{multicols}{2}

\infrule[Sel2]
{\Gamma \ts x: (L:T..\top)}
{\Gamma \ts T \sub x.L}

\infrule[Sel1]
{\Gamma \ts x: (L:\bot..T)}
{\Gamma \ts x.L \sub T}

\end{multicols}

\bigskip
Recursive self types

\infrule[BindX]
{\Gamma,z:T_1^z \ts T_1^z \sub T_2^z}
{\Gamma \ts \{ z \Rightarrow T_1^z \} \sub \{z \Rightarrow T_2^z \}}

\infrule[Bind1]
{\Gamma,z:T_1^z \ts T_1^z \sub T_2}
{\Gamma \ts \{ z \Rightarrow T_1^z \} \sub T_2}

\bigskip
Transitivity

\infrule[Trans]
{\Gamma \ts T_1 \sub T_2 \spcomma T_2 \sub T_3}
{\Gamma \ts T_1 \sub T_3}

\columnbreak

\judgement{Type assignment}{\fbox{$\Gamma \ts t : T$}}

\typicallabel{}
\smallrulenames

Variables, self packing/unpacking

\infrule[Var]
{\Gamma(x) = T}
{\Gamma \ts x : T}

\setlength{\columnseprule}{0pt}
\begin{multicols}{2}

\infrule[VarPack]
{\Gamma \ts x : T^x}
{\Gamma \ts x : \{ z \Rightarrow T^z \}}

\infrule[VarUnpack]
{\Gamma_{[x]} \ts x : \{ z \Rightarrow T^z \}}
{\Gamma \ts x : T^x}

\end{multicols}

\bigskip
Subsumption

\infrule[Sub]
{\Gamma \ts t : T_1 \spcomma T_1 \sub T_2}
{\Gamma \ts t : T_2}

\bigskip
Field selection, method invocation

\infrule[TFld]
{\Gamma \ts t: (l:T)}
{\Gamma \ts t.l : T}

\infrule[TFunVar]
{\Gamma \ts t: (m(x:T_1):T_2^x) \spcomma y: T_1}
{\Gamma \ts t.m(y): T_2^y}

\infrule[TFun]
{\Gamma \ts t: (m(x:T_1):T_2) \spcomma t_2: T_1}
{\Gamma \ts t.m(t_2): T_2}

\bigskip
Object creation and member initialization

\infrule[TNew]
{\text{\scriptsize(labels disjoint)}\\\Gamma, x:T_1^x \tand \ldots \tand T_n^x \ts d_i: T_i^x\gap\forall i, 1\leq i \leq n}
{\Gamma \ts \{ x \Rightarrow d_1 \ldots d_n \}: \{ z \Rightarrow T_1^z \tand \ldots \tand T_n^z \}}

\infrule[DFld]
{\Gamma \ts t: T}
{\Gamma \ts (l = t): (l:T)}

\infrule[DMem]
{\Gamma \ts T \sub T}
{\Gamma \ts (L = T): (L: T..T)}

\infrule[DFun]
{\Gamma,x:T_1 \ts t: T_2}
{\Gamma \ts (m(x) = t): (m(x:T_1):T_2)}

\end{multicols}
}

Figure~\ref{fig:dot1} shows the syntax and static semantics of 
the DOT calculus we study.
For readability, we omit well-formedness requirements from the rules,
and assume all types to be syntactically well-formed in the
given environment. We write $T^x$ when $x$ is free in $T$.

Compared to the original DOT proposal \cite{dotfool},
which used several auxiliary judgments such as membership and expansion,
the presentation here is vastly simplified, and uses only subtyping 
to access function and type members. Compared to $\mu$DOT
\cite{DBLP:conf/oopsla/AminRO14}, the calculus is much more expressive,
and includes key features like intersection and union types, which
are absent in $\mu$DOT.

The Scala syntax used above maps to the formal notation in
a straighforward way:
\begin{lstlisting}
          { type T = Elem }  $\rightarrow$  $T:\text{Elem}..\text{Elem}$
       { type T >: S <: U }  $\rightarrow$  $T:S..U$
        { def m(x: T) = t }  $\rightarrow$  $m(x)=t$
              A & B,  A | B  $\rightarrow$  $A \tand B, \ A \tor B$
\end{lstlisting}

Intersection and union types, along with the $\bot$ and $\top$ types, 
provide a full subtyping lattice.

In subtyping, members of the same label and kind can be compared. The
field type, type member upper bound, and method result type are
covariant while the type member lower bound and the method parameter
type are contravariant -- as is standard. We allow some dependence
between the parameter type and the return type of a method, when the
argument is a variable. This is another difference to previous
versions of DOT \cite{dotfool,DBLP:conf/oopsla/AminRO14}, which 
did not have such dependent method types.

If a variable $x$ can be assigned a type member $L$, then the type
selection $x.L$ is valid, and can be compared with any upper bound
when it is on the left, and with any lower bound when it is on the
right. Furthermore, a type selection is reflexively a subtype of
itself. This rule is explicit, so that even abstract type members can
be compared to themselves.

Finally, recursive self types can be compared to each other as intuitively
expected. They can also be dropped if the self identifier is not used. 
During type assignment, the rules for variable packing and 
unpacking serve as introduction and elimination rules, enabling 
recursive types to be compared to other types as well.
Since subtype comparisons may introduce temporary bindings that may
need to be unpacked, the unpack rule comes with a syntactic restriction
to ensure termination in the proofs (Section~\ref{sec:oncemore}).
In general, environments have the form $\Gamma = \seq{y:T},\seq{z:T}$, 
with term bindings followed by bindings introduced by subtype comparisons.
The notation $\Gamma_{[x]}$ in the unpacking rule signifies that 
all $z\!:\!T$ bindings to the right of $x$ are dropped from $\Gamma$. 
While this restriction is necessary for the proofs,
it does not seem to limit expressiveness of the type system in any
significant way. 
Outside of subtyping comparisons involving binders, $\Gamma_{[x]}$ 
is just $\Gamma$. Thus, the restriction is irrelevant for 
variable unpacking in normal type assignment.

In the version presented here, we make the transitivity rule explicit,
although, as we will see in Section~\ref{sec:main3}, we will sometimes
need to ensure that we can eliminate uses of this rule from
subtyping derivations so that the last rule is a structural one. 

The aim of DOT is to be a simple, foundational calculus in the
spirit of FJ~\cite{DBLP:journals/toplas/IgarashiPW01}. The aim is
not to commit to specific decisions for nonessential things. 
Hence, implementation inheritance, mixin strategy, and prototype vs class
dispatch are not considered.

\section{Static Properties of DOT}\label{sec:main3}

Having introduced the syntax and static semantics of DOT, we turn to its 
metatheoretic properties. Our main focus of interest will be type safety:
establishing that well-typed DOT programs do not go wrong.  Of course,
type safety is only meaningful with respect to a dynamic semantics, which 
we will discuss in detail in Sections~\ref{sec:defint} and \ref{sec:fsub}.
Here, we briefly touch some general static properties of DOT and
then discuss specific properties of the subtyping relation, which 
(or their absence!) makes proving type safety a challenge.

\paragraph{Decidability}
Type assignment and subtyping are undecidable in DOT. This follows directly
from the fact that DOT can encode F$_{\sub}$, and that these properties are
undecidable there.

\paragraph{Type Inference}
DOT has no principal types and no global Hindley-Milner style type inference
procedure. But as in Scala, local type inference based on subtype constraint 
solving \cite{DBLP:journals/toplas/PierceT00,DBLP:conf/popl/OderskyL96} is 
possible, and in fact easier than in Scala due to the existence of 
universal greatest lower bounds and least upper bounds through intersection and 
union types. For example, in Scala, the least upper bound of the two types
@C@ and @D@ is approximated by an infinite sequence:
\begin{lstlisting}
trait A { type T <: A }
trait B { type T <: B }
trait C extends A with B { type T <: C }
trait D extends A with B { type T <: D }
lub(C, D) ~ A with B { type T <: A with B { type T <: ... } }
\end{lstlisting}
DOT's intersection and union types remedy this brittleness.

While the term syntax and type assignment 
given in Figure~\ref{fig:dot1} is presented in Curry-style, without explicit 
type annotations except for type member initializations, a Church-style version 
with types on method arguments (as required for local type inference) is 
possible just as well.

\subsection{Properties of Subtyping}
The relevant static properties we are interested in with regard
to type safety are transitivity, narrowing, and inversion
of subtyping and type assignment. They are defined as
follows.

Inversion of subtyping (example: functions):
\infrule[\textsc{InvFun}]
{\Gamma \ts m(x:S_1):U_1^x \sub m(x:S_2):U_2^x}
{\Gamma \ts S_2 \sub S_1\gap\Gamma,x:S_2 \ts U_1^x \sub U_2^x}

Transitivity of subtyping:
\infrule[\textsc{Trans}]
{\Gamma \ts T_1 \sub T_2 \spcomma T_2 \sub T_3}
{\Gamma \ts T_1 \sub T_3}

Narrowing:
\infrule[\textsc{Narrow}]
{\Gamma_1 \ts T_1 \sub T_2 \gap \Gamma_2 \ts T_3 \sub T_4\\ \Gamma_1=\Gamma_2(x\rightarrow T_1) \gap \Gamma_2(x)=T_2}
{\Gamma_1 \ts T_3 \sub T_4}

On a high-level, the basic challenge for type safety is to establish that
some value that e.g.\ has a function type actually is a function, with arguments
and result corresponding to the given type. This is commonly known as the
\emph{canonical forms} property. Inversion of subtyping is required to extract 
the argument and result types from a given subtype relation between two 
function types, in particular to derive
$$T_2 <: T_1 \text{ and } U_1 <: U_2$$ from $$m(x:T_1):U_1 <: m(x:T_2):U_2$$
when relating method types from a call site and the definition
site.

Transitivity is required to collapse multiple subsumption 
steps that may have been used in type assignment. Narrowing can be seen as 
an instance of the Liskov substitution principle, preserving subtyping if 
a binding in the environment is replaced with a subtype.

Unfortunately, as we will show next, these properties do \emph{not}
hold simultaneously in DOT, at least not in their full generality.

\subsection{Inversion, Transitivity and Narrowing}\label{sec:2trans}

First of all, let us take note that these properties are
mutually dependent. In Figure~\ref{fig:dot1}, we have
included $(\textsc{Trans})$ as an axiom. If we drop this axiom, the
rules become syntax directed and we obtain rules like 
$(\textsc{InvFun})$ by direct inversion of the
corresponding typing derivation. 
But then, we would need to prove $(\textsc{Trans})$ as a lemma.

Transitivity and narrowing are also mutually dependent.
Transitivity requires narrowing in the following case:
$$\{ z \Rightarrow T_1 \} <: \{ z \Rightarrow T_2 \} <: \{ z \Rightarrow T_3 \}$$

By inversion we obtain
$$z:T_1 \ts T_1 <: T_2  \gap z:T_2 \ts T_2 <: T_3$$

and we narrow the right-hand derivation to $z:T_1 \ts T_2 <: T_3$
before we apply transitivity inductively to obtain $z:T_1 \ts T_1 <: T_3$.

Narrowing depends on transitivity in the case for type selections
$$x.L <: T \text{ or its counterpart } T <: x.L$$

Assume that we want to narrow $x$'s binding from 
$T_2$ in $\Gamma_2$ to $T_1$ in $\Gamma_1$, with $\Gamma_1 \ts T_1 <: T_2$. 
By inversion we obtain
$$x: (L: \bot..T)$$
and, disregarding rules ($\textsc{VarPack}$) and ($\textsc{VarUnpack}$) we can
deconstruct this assignment as 
$$\Gamma_2(x) = T_2  \gap  \Gamma_2 \ts T_2 <: (L: \bot..T).$$

We first apply narrowing inductively and then use
transitivity to derive the new binding
$$\Gamma_1(x) = T_1  \gap  \Gamma_1 \ts T_1 <: T_2 <: (L: \bot..T).$$

On first glance, the situation appears to be similar
to simpler calculi like F$_{\sub}$, for which the transitivity
rule can be shown to be admissible, i.e.\ implied by other
subtyping rules and hence proved as a lemma and dropped
from the definition of the subtyping relation.
Unfortunately this is not the case in DOT.

\subsection{Good Bounds, Bad Bounds}\label{sec:goodbounds}

The transitivity axiom (or subsumption step in type assignment) 
is essential and cannot be dropped. Let us go through and see 
why we cannot prove transitivity directly.

First of all, observe that transitivity can only hold if all types 
in the environment have `good bounds', i.e.\ only members
where the lower bound is a subtype of the upper bound. Here is
a counterexample. Assume a binding with `bad' bounds like $\texttt{Int}..\texttt{String}$. 
Then the following subtype relation holds via transitivity
$$x:\{A=\texttt{Int}..\texttt{String}\} \ts \texttt{Int} <: x.A <: \texttt{String}$$
but @Int@ is not a subtype of @String@. Of course core DOT does
not have @Int@ and @String@ types, but any other incompatible 
types can be taken as well.

But even if we take good bounds as a precondition, we cannot show
$$\{ z \Rightarrow T_1 \} <: \{ z \Rightarrow T_2 \} <: \{ z \Rightarrow T_3 \}$$
because we would need to use $x:T_1$ as extended environment for the
inductive call, but we do not know that @T1@ has indeed good bounds.

Of course we could modify the $\{ z \Rightarrow T_1 \} <: \{ z \Rightarrow T_2 \}$ rule to 
require $T_1$ to have good bounds. But then this evidence would need 
to be narrowed, which unfortunately is not possible. Again, here is a 
counterexample:
$$x:\{A:\bot..\top\} \ts x.A \tand \{B=\texttt{Int}\}$$
This type has good bounds, but when narrowing $x$ to the smaller 
type $\{A=\{B=\texttt{String}\}\}$
(which also has good bounds), its bounds become contradictory.

In summary, even if we assume good bounds in the environment, and 
strengthen our typing rules so that only types with good bounds
are added to the environment, we may still end up with bad bounds
due to narrowing and intersection types. This refutes one
conjecture about possible proof avenues from earlier work on DOT
\cite{DBLP:conf/oopsla/AminRO14}.

\subsection{No (Simple) Substitution Lemma}\label{sec:nosubst}

To finish off this section, we observe how typing is not preserved 
by straightforward substitution in DOT, for the simple existence
of path-dependent types:
$$\Gamma,x:\{z \Rightarrow L: S^z..U^z \tand l: T^z\} \ts t: x.L$$
First, $t$ might access field $x.l$ and therefore require
assigning type $(l:T^x)$ to $x$. However, the self bind layer
can only be removed if an object is accessed through a variable,
otherwise we would not be able to substitute away the self identifier $z$.
Second, we cannot derive $\Gamma \ts t: x.L$ with $x$
removed from the environment, but we also cannot replace $x$ in $x.L$
with a term that is not an identifier.

Of course we can think about clever ways to circumvent these 
issues but the details become intricate quickly, with relating
variables before and after reduction becoming a key 
concern \cite{dotfool}. Moreover, the issues with narrowing
and bad bounds remain, and if we assign type @String@ to
an @Int@ value through a transitivity step, we have an actual
type safety violation.

While ultimately, suitable substitution strategies have been
found (see Section~\ref{sec:smallstep}), these results have been 
enabled by changing the perspective, and taking a very high-level
execution model based on definitional interpreters as a
starting point.

Intuitively, all the questions related to bad bounds have a 
simple answer: we ensure good bounds at object creation time, 
\emph{so why do we even need to worry about all the intermediate 
steps in such a fine-grained way?}

\section{Definitional Interpreters for Type Soundness}\label{sec:defint}

Today, the dominant method for proving soundness of a type system is the 
syntactic approach of Wright and Felleisen \cite{DBLP:journals/iandc/WrightF94}. 
Its key components are the progress and preservation lemmas with respect 
to a small-step operational semantics based on term rewriting.
While this syntactic approach has a lot of benefits, as described in great
detail in the original 1994 paper \cite{DBLP:journals/iandc/WrightF94}, 
there are also some drawbacks.
An important one is that reduction semantics often pose a question of 
adequacy: realistic language implementations do not proceed by rewriting, 
so if the aim is to model an existing language, at least an informal argument 
needs to be made that the given reduction relation faithfully implements 
the intended semantics. Furthermore, few realistic languages actually enjoy the 
subject reduction property. If simple substitution does not hold, the syntactic
approach is more difficult to apply and requires stepping into richer 
languages in ways that are often non-obvious. Again, care must be taken
that these richer languages are self-contained and match the original
intention.

We have already seen that naive substitution does not preserve types
in DOT (Section~\ref{sec:nosubst}). In fact, this is also true for
many other languages, or language features. 

\paragraph{Example 1: Return statements}
Consider a simple program in a language with @return@ statements:
\begin{lstlisting}
  def fun(c) = if (c) return x; y
  fun(true)
\end{lstlisting}
Taking a straightforward small-step execution strategy, this
program will reduce to:
\begin{lstlisting}
  $\reduces$ if (true) return x; y
\end{lstlisting}
But now the @return@ has become unbound. We need to augment the 
language and reduce to an auxiliary construct like this:
\begin{lstlisting}  
  $\reduces$ scope { if (true) return x; y } 
\end{lstlisting}
This means that we need to work with a richer language than
we had originally intended, with additional syntax, typing, and 
reduction rules like the following:
\begin{lstlisting}  
  scope E[ return v ] $\reduces$ v  $~~~~$    scope v $\reduces$ v
\end{lstlisting}

\paragraph{Example 2: Private members}
As another example, consider an object-oriented language with
access modifiers.
\begin{lstlisting}
  class Foo { private val data = 1;  def foo(x) = x * this.data }
\end{lstlisting}
Starting with a term
\begin{lstlisting}
  val a = new Foo; a.foo(7) / S
\end{lstlisting}
where @S@ denotes a store, small-step reduction steps may lead to:
\begin{lstlisting}
  $\reduces$ l0.foo(7)                / S, (l0 -> Foo(data=1))
  $\reduces$ x * l0.data              / S, (l0 -> Foo(data=1))
\end{lstlisting}
But now there is a reference to private field @data@ outside the scope of class @Foo@.

We need a special rule to ignore access modifiers for `runtime' objects in the store, 
versus other expresssions that happen to have type @Foo@. We still want to disallow
@a.data@ if @a@ is a normal variable reference or some other expression.

\paragraph{Example 3: Method overloading}
Looking at a realistic language, many type preservation issues are documented 
in the context of Java, which were discussed at length on the Types mailing 
list, back in the time when Java's type system was an object of study
\footnote{Subject Reduction fails in Java: \url{http://www.seas.upenn.edu/~sweirich/types/archive/1997-98/msg00452.html}}.

Most of these examples relate to static method overloading, and to
Java's conditional expressions @c ? a : b@, which require @a@ and @b@
to be in a subtype relationship because Java does not have least
upper bounds. It is worth noting that these counterexamples to
preservation are not actual type safety violations.

\subsection{Alternative Semantic Models}

So what can we do if our object of study does not naturally fit
a rewriting model of execution? Of course one option is to
make it fit (perhaps with force), but an easier path may be to pick a 
different semantic model. Before the syntactic approach became popular, 
denotational semantics \cite{DBLP:conf/icalp/Scott82}, Plotkin's structural 
operational semantics \cite{DBLP:journals/jlp/Plotkin04a} and Kahn's natural semantics 
(or `big-step' semantics) \cite{DBLP:conf/stacs/Kahn87} were the tools of the trade. 

Big-step semantics in particular has the benefit of being more
`high-level', in the sense of being closer to actual language
implementations. Their downside for soundness proofs is
that failure cases and nontermination are not easily
distinguished. This often requires tediously enumerating
all possible failure cases, which may cause a blow-up in the 
required rules and proof cases. Moreover, in the history of 
big-step proofs, advanced language features such as recursive 
references have required specific proof techniques (e.g.\ coinduction) 
\cite{Tofte88operationalsemantics} 
which made it hard to compose proofs for different language features. 
In general, polymorphic type systems pose difficulties, because
type variables need to be related across different contexts.

But the circumstances have changed since 1994. Today,
most formal work is done in proof assistants
such as Coq, and no longer with pencil and paper. This means
that we can use software implementation techniques like monads 
(which, ironically were developed in the context of denotational
semantics \cite{DBLP:journals/iandc/Moggi91}) to handle the complexity of failure cases. 
Moreover, using simple but clever inductive techniques such
as step counters we can avoid the need for coinduction and
other complicated techniques in many cases. 

In the following, we present our approach to type soundness
proofs with definitional interpreters in the style
of Reynolds \cite{DBLP:journals/lisp/Reynolds98a}: 
high-level evaluators implemented in 
a (total) functional language. In a functional
system such as Coq or Agda, implementing evaluators is more
natural than implementing relational transition systems, 
with the added benefit of always having a directly 
executable model of the language at hand.

We present a fully mechanized type safety proof for System F$_{\sub}$,
which we gradually extend to DOT.
This proof of F$_{\sub}$ alone is significant, because it shows that 
indeed type safety proofs for polymorphic type systems can be nice 
and simple using big-step techniques, which correspond closely to 
actual language implementations. Deriving DOT as an extension of
F$_{\sub}$ has also lead to important insights that were not
apparent in previous models of the calculus. In particular, the
intermediate systems like D$_{\sub}$ in Section~\ref{sec:dsub} inhabit 
interesting points in the design space of dependently typed calculi 
between F$_{\sub}$ and full DOT.

\subsection{Simply Typed Lambda Calculus: Siek's 3 Easy Lemmas}
\label{sec:stlc}

We build our exposition on Siek's type safety proof for a dialect
of simply typed lambda calculus (STLC) \cite{siek3lemmas}, 
which in turn takes 
inspiration from Ernst, Ostermann and Cook's semantics in their 
formalization of virtual classes \cite{vc}.

The starting point is a fairly standard definitional interpreter 
for STLC, shown in Figure~\ref{fig:stlc} together with the
STLC syntax and typing rules. We opt to show the interpreter
in actual Coq syntax, but stick to formal notation for the
language definition and typing rules. The interpreter consists 
of three functions: one for primitive operations (which we elide), 
one for variable lookups, and one main evaluation function @eval@,
which ties everything together. Instead of working exclusively 
on terms, as a reduction semantics would do, the interpreter
maps terms to a separate domain of values @v@. Values include 
primitives, and closures, which pair a term with an environment.

\begin{figure}[h!]
\begin{framed}
\judgement{Syntax}{}

\medskip

$\ba{lll}
  T &::=& B \ |\ T \rightarrow T\\
  t &::=& c \ |\ x \ |\  \lambda x:T.t \ |\ t\ t\\
  v &::=& c \ |\ \clos{H,\lambda x:T.t}\\
  r &::=& @Timeout@ \ |\ @Done@\ (@Error@\ |\ @Val@\ v)\\
  \Gamma &::=& \emptyset \ |\ \Gamma,x:T\\
  H &::=& \emptyset \ |\ H,x:v
\ea$

\medskip

\judgement{Type assignment}{\fbox{$\Gamma \ts t : T$}}

  \infax[]{\Gamma \ts c: B}

  \infrule[]{x:T \in \Gamma}
  {\Gamma \ts x: T}

  \infrule[]{\Gamma,x:T_1 \ts t: T_2}
  {\Gamma \ts \lambda x:T_1.t: T_1 \rightarrow T_2}

  \infrule[]{\Gamma \ts t_1: T_1 \rightarrow T_2  \spcomma t_2: T_1}
  {\Gamma \ts t_1 t_2 : T_2}

\judgement{Consistent environments}{\fbox{$\Gamma \tS H$}}

  \infax[]{\emptyset \tS \emptyset}

  \infrule[]{\Gamma \tS H \gap v: T}
  {\Gamma,x:T \tS H,x:v}

\judgement{Value type assignment}{\fbox{$H \ts v : T$}}

  \infax[]{c: B}

  \infrule{\Gamma \tS H  \gap \Gamma,x:T_1 \ts t: T_2}
  {\clos{H,\lambda x:T_1.t}: T_1 \rightarrow T_2}

\judgement{Definitional Interpreter}{}

\begin{lstlisting}[keywords={}]
(* Coq data types and auxiliary functions elided *)
Fixpoint eval(n: nat)(env: venv)(t: tm){struct n}: 
option (option vl) :=
  DO n1 <== FUEL n;                            (* totality    *)
  match t with
    | tcst c      => DONE VAL (vcst c)         (* constant    *)
    | tvar x      => DONE (lookup x env)       (* variable    *)
    | tabs x ey   => DONE VAL (vabs env x ey)  (* lambda      *)
    | tapp ef ex  =>                           (* application *)
      DO vf <== eval n1 env ef;
        DO vx <==  eval n1 env ex;
        match vf with
          | (vabs env2 x ey) =>
            eval n1 ((x,vx)::env2) ey
          | _ => ERROR
        end
  end.
\end{lstlisting}
\caption{\label{fig:stlc}STLC: Syntax and Semantics}
\end{framed}
\vspace{1.4cm}
\end{figure}

\paragraph{Notions of Type Soundness}
What does it mean for a language to be type safe? We follow 
Wright and Felleisen \cite{DBLP:journals/iandc/WrightF94} 
in viewing a static type system as a filter that selects well-typed 
programs from a larger universe of untyped programs. In their
definition of type soundness, a partial function @evalp@ defines 
the semantics of untyped programs, returning @Error@ if the  evaluation
encounters a type error, or any other answer for a well typed
result. We assume here that the result in this case will be 
@Val v@, for some value @v@. For evaluations that do not 
terminate, @evalp@ is undefined.

The simplest soundness property states that well-typed programs 
do not go wrong.

Weak soundness:
\infrule[]
{\emptyset \ts e: T}
{\text{evalp } e\ \ne \text{ Error}}

A stronger soundness property states that if the evaluation
terminates, the result value must have the same type as the
program expression, assuming that values are classified
by types as well.

Strong soundness:
\infrule[]
{\emptyset \ts e: T \gap \text{evalp } e = r}
{r = \text{ Val }v \gap v: T}

In our case, assigning types to values is achieved
by the rules in the lower half of Figure~\ref{fig:stlc}.

\paragraph{Partiality Fuel}
To reason about the behavior of our interpreter, and
to implement it in Coq in the first place, we had to
get a handle on potential nontermination, and make
the interpreter a total function. Again we follow
\cite{siek3lemmas} by first making all error cases 
explicit by wrapping the result of each operation in 
an @option@ data type with alternatives @Val v@ and @Error@. 
This leaves us with possible nontermination. To model
nontermination, the interpreter is parameterized over
a step-index or `fuel value' @n@, which bounds the 
amount of work the interpreter is allowed to do. 
If it runs out of fuel, the interpreter returns 
@Timeout@, otherwise @Done r@, where @r@ is the
option type introduced above. 

It is convenient to treat this type of answers as
a (layered) monad and write the interpreter in monadic
@do@ notation (as done in Figure~\ref{fig:stlc}).
The @FUEL@ operation in the first line desugars
to a simple non-zero check:
\begin{lstlisting}[keywords={}]
    match n with 
      | z => TIMEOUT 
      | S n1 => ...
    end
\end{lstlisting}
There are other ways to define monads that encode
partiality (see e.g. \cite{DBLP:conf/icfp/Danielsson12}
for a treatment that involves a coinductively defined 
partiality monad), but this simple method has the benefit 
of enabling easy inductive proofs about all executions 
of the interpreter, by performing a simple induction 
over $n$. If a fact is proved for all executions of length 
$n$, for all $n$, then it must hold for all finite executions.
Specifically, infinite executions are by definition not 
`stuck', so they cannot violate type soundness.

\paragraph{Proof Structure}
For the type safety proof, the `three easy lemmas' \cite{siek3lemmas}
are as follows. There is one lemma per function 
of the interpreter. The first one shows that well-typed 
primitive operations are not stuck and preserve types. 
We are omitting primitive operations for simplicity,
so we skip this lemma.
The second one shows that well-typed environment 
lookups are not stuck and preserve types.
\infrule[]
{\Gamma \tS H \gap \text{lookup } x \ \Gamma = \text{ Some } T}
{\text{lookup } x \ H = \text{ Val } v \gap v: T}
This lemma is proved by structural induction
over the environment and case distinction whether the 
lookup succeeds.

The third lemma shows that, for all $n$, if the interpreter
returns a result that is not a timeout, the result is
also not stuck, and it is well typed.
\infrule[]
{\Gamma \ts e: T \gap \Gamma \tS H \gap \text{eval } n\ H\ e = \text{Done } r}
{r = \text{Val } v \gap v: T \gap}
The proof is by induction on $n$, and case analysis
on the term $e$.

It is easy to see that this lemma corresponds to the strong 
soundness notion above. In fact, we can define a partial function
$\text{evalP } e$ to probe $\text{eval } n\ \emptyset\ e$ for all $n = 0,1,2,...$
and return the first non-timeout result, if one exists. Restricting
to the empty environment then yields \emph{exactly} Wright and Felleisen's
definition of strong soundness:
\infrule[]
{\emptyset \ts e: T \gap \text{evalp } e = r}
{r = \text{ Val }v \gap v: T}

Using classical reasoning we can conclude that either 
evaluation diverges (times out for all @n@), or
there exists an @n@ for which the result is 
well typed.

\section{Type Soundness for System F$_{<:}$}\label{sec:fsub}

We now turn our attention to System F$_{\sub}$ \cite{DBLP:journals/iandc/CardelliMMS94}, 
moving beyond type systems that have been previously formalized with 
definitional interpreters. The syntax and static 
typing rules of F$_{\sub}$ are defined in Figure~\ref{fig:fsub1}. 

\begin{figure}[h!]
\begin{framed}
\judgement{Syntax}{}

\medskip

$\ba{lll}
  X &::=& Y \ |\ Z\\
  T &::=& X \ |\ \Top \ |\ T \rightarrow T \ |\ \forall Z<:T.T^Z\\
  t &::=& x \ |\ \lambda x:T.t \ |\ \Lambda Y<:T.t \ |\ t\ t \ |\ t\ [T]\\
  \Gamma &::=& \emptyset \ |\ \Gamma,x:T \ |\ \Gamma,X<:T
\ea$

\medskip

\judgement{Subtyping}{\fbox{$\Gamma \ts S \sub U$}}

  \infax[]{\Gamma \ts S \sub \Top}

  \infax[]{\Gamma \ts X \sub X}

  \infrule[]{\Gamma \ni X<:U \gap \Gamma \ts U \sub T}
  {\Gamma \ts X \sub T}

  \infrule[]{\Gamma \ts T_1 \sub S_1 \spcomma S_2 \sub T_2}
  {\Gamma \ts S_1 \rightarrow S_2 \sub T_1 \rightarrow T_2}

  \infrule[]{\Gamma \ts T_1 \sub S_1\\\Gamma,Z<:T_1 \ts S_2^Z \sub T_2^Z}
  {\Gamma \ts \forall Z<:S_1.S_2^Z \sub \forall Z<:T_1.T_2^Z}

\judgement{Type assignment}{\fbox{$\Gamma \ts t : T$}}

  \infrule[]{\Gamma \ni x:T}
  {\Gamma \ts x: T}

  \infrule[]{\Gamma,x:T_1 \ts t_2: T_2}
  {\Gamma \ts \lambda x:T_1.t_2: T_1 \rightarrow T_2}

  \infrule[]{\Gamma \ts t_1: T_1 \rightarrow T_2 \spcomma t_2: T_1}
  {\Gamma \ts t_1 t_2: T_2}

  \infrule[]{\Gamma,Y<:T_1 \ts t_2: T_2^Y}
  {\Gamma \ts \Lambda Y<:T_1.t_2: \forall Z<:T_1.T_2^Z}

  \infrule[]{\Gamma \ts t_1: \forall Z<:T_{11}.T_{12}^Z \spcomma T_2 \sub T_{11}}
  {\Gamma \ts t_1 [T_2]: T_{12}^{T_2}}

  \infrule[]{\Gamma \ts t: S \spcomma S \sub T}
  {\Gamma \ts t: T}
          
\caption{\label{fig:fsub1}F$_{\sub}$: syntax and static semantics}
\end{framed}
\end{figure}

In addition to STLC, we have type abstraction and type application, 
and subtyping with upper bounds. The calculus is more expressive 
than STLC and more interesting from a formalization perspective, in particular 
because it contains type variables. 
These are bound in the environment, which means that we need to consider 
types in relation to the environment they were defined in.

What would be a suitable runtime semantics for passing type arguments to
type abstractions?
The usual small-step semantics uses substitution to eliminate type arguments:
$$  (\Lambda X<:T.t) [T] \longrightarrow t[T/X]$$
We could do the same in our definitional interpreter: 
\begin{lstlisting}[keywords={}]
    | ttapp ef T =>
      DO vf <== eval n1 env ef;
        match vf with
          | (vtabs env2 x ey) => eval n1 env2 (substitute ey x T)
          | _ => ERROR
        end
\end{lstlisting} 
But then, the interpreter would have to modify program terms at runtime. This would be odd for an interpreter, which is meant to be simple. It would also complicate formal reasoning, since we would still need a substitution lemma like in small step semantics.

\subsection{Types in Environments}

\begin{figure}[h!]
\begin{framed}
\judgement{Syntax}{}

\medskip

$\ba{lll}
  v &::=& \clos{H,\lambda x:T.t} \ |\ \clos{H,\Lambda Y<:T.t}\\
  H &::=& \emptyset \ |\ H,x:v \ |\  H,Y=\clos{H,T}\\
  J &::=& \emptyset \ |\ J,Z<:\clos{H,T}
\ea$

\medskip

\judgement{Runtime subtyping}{\fbox{$J \ts H_1\ T_1 \sub H_2\ T_2$}}

  \infax[]{J \ts H_1\ T \sub H_2\ \Top}

  \infrule[]{J \ts H_2\ T_1 \sub H_1\ S_1 \gap J \ts H_1\ S_2 \sub H_2\ T_2}
  {J \ts H_1\ S_1 \rightarrow S_2\; \sub\; H_2\ T_1 \rightarrow T_2}

  \infrule[]{J \ts H_2\ T_1 \sub H_1\ S_1\\
  J,Z\!<:\!\clos{H_2,T_1} \ts H_1\ S_2^Z\; \sub\; H_2\ T_2^Z}
  {J \ts H_1\ \forall Z\!<:\!S_1.S_2^Z\; \sub\; H_2\ \forall Z\!<:\!T_1.T_2^Z}

Abstract type variables

  \infax[]{J \ts H_1\ Z \sub H_2\ Z}

  \infrule[]{J \ni Z<:\clos{H,U} \gap J \ts H\ U \sub H_2\ T}
  {J \ts H_1\ Z \sub H_2\ T}

Concrete type variables

  \infrule[]{H_1 \ni Y_1=\clos{H,T} \gap H_2 \ni Y_2=\clos{H,T}}
  {J \ts H_1\ Y_1 \sub H_2\ Y_2}

  \infrule[]{H_1 \ni Y=\clos{H,U} \gap J \ts H\ U \sub H_2\ T}
  {J \ts H_1\ Y \sub H_2\ T}

  \infrule[]{J \ts H_1\ T \sub H\ L \gap H_2 \ni Y=\clos{H,L}}
  {J \ts H_1\ T \sub H_2\ Y}

Transitivity

  \infrule[]{J \ts H_1\ T1 \sub H_2\ T2 \gap H_2\ T_2 <: H_3\ T_3}
  {J \ts H_1\ T_1 \sub H_3\ T_3}

\judgement{Consistent environments}{\fbox{$\Gamma \tS H\ J$}}

  \infax[]{\emptyset \tS \emptyset\ \emptyset}

  \infrule[]{\Gamma \tS H\ J \gap  H \ts v: T}
  {\Gamma,x:T \tS (H,x:v)\ J}

  \infrule{\Gamma \tS H\ J \gap J \ts H_1\ T_1 \sub H\ T}
  {\Gamma,Y<:T \tS (H,Y=\clos{H_1,T_1})\ J}

  \infrule{\Gamma \tS H\ J \gap J \ts H_1\ T_1 \sub H\ T}
  {\Gamma,Z<:T \tS H\ (J,Z<:\clos{H_1,T_1})}

\judgement{Value type assignment}{\fbox{$H \ts v : T$}}

  \infrule[]{\Gamma \tS H\ \emptyset \gap \Gamma,x:T_1 \ts t: T_2}
  {H \ts \clos{H,\lambda x:T_1.t}: T_1 \rightarrow T_2}

  \infrule[]{\Gamma \tS H\ \emptyset \gap \Gamma,Y<:T_1 \ts t: T_2^Y}
  {H \ts \clos{H,\Lambda Y<:T_1.t}: \forall Z<:T_1.T_2^Z}

  \infrule[]{H_1 \ts v: T_1  \gap \emptyset \ts H_1\ T_1 \sub H_2\ T_2}
  {H_2 \ts v: T_2}
          
\caption{\label{fig:fsub2}F$_{\sub}$: dynamic typing}
\end{framed}
\end{figure}

A better idea, more consistent with an environment-passing 
interpreter, is to put the type argument into the runtime 
environment as well:
\begin{lstlisting}[keywords={}]
          | (vtabs env2 x ey) => eval n1 ((x,T)::env2) ey
\end{lstlisting} 
However, this leads to a problem: the type @T@ may refer to 
other type variables that are bound in the current environment 
at the call site. We could potentially resolve all the references,
and substitute their occurrences in the type, but this will no
longer work if types are recursive. Instead, we pass the
caller environment along with the type. 
\begin{lstlisting}[keywords={}]
          | (vtabs env2 x ey) => eval n1 ((x,vty env T)::env2) ey
\end{lstlisting} 
In effect, type arguments become very similar to function closures, 
in that they close over their defining environment.
Intuitively, this makes a lot of sense, and is consistent with the
overall workings of our interpreter.

But now we need a subtyping judgement that takes the respective
environments into account when comparing two types at runtime.
This is precisely what we do. The runtime typing rules are
shown in Figure~\ref{fig:fsub2}. We will explain the role of
the $J$ environments shortly, in Section~\ref{sec:6hyp} below, 
but let us first take note that the runtime subtyping judgement 
takes the form 
$$J \ts H_1\ T_1 <: H_2\ T_2,$$
pairing each type $T$ with a corresponding runtime
environment $H$.

Note further that the rules labeled `concrete type variables'
are entirely structural: different type variables $Y_1$ and
$Y_2$ are treated equal if they map to the same $H\ T$ pair. 
In contrast to the surface syntax of
F$_{\sub}$, there is not only a rule for $Y <: T$ but also
a symmetric one for $T <: Y$, i.e.\ with a type variable on
the right hand side. This symmetry is necessary to model
runtime type equality through subtyping, which gives us a
handle on those cases where a small-step semantics would 
rely on type substitution.

Another way to look at this is that the dynamic subtyping 
relation removes abstraction barriers (nominal variables,
only one sided comparison with other types) that were put 
in place by the static subtyping relation.

We further note in passing that subtyping transitivity becomes more
difficult to establish than for static F$_{\sub}$ subtyping, 
because we now may have type variables in the middle of a 
chain $T_1 <: Y <: T_2$, which should contract to $T_1 <: T_2$.
We will get back to this question and related ones in 
Section~\ref{sec:6trans}.

\subsection{Abstract Types}\label{sec:6hyp}

So far we have seen how to handle types that correspond to existing type objects.
We now turn to the rule that compares $\forall$ types, and which introduces new 
type bindings when comparing two type abstractions. 

How can we support this rule at runtime? We cannot quite use only the facilities
discussed so far, because this would require us to `invent' new hypothetical 
objects $\clos{H,T}$, which are not actually created during execution, and insert them into another runtime environment.
Furthermore, to support transitivity of the $\forall$ rule we need narrowing, and we would not want to think about potentially replacing actual values in a runtime environment with something else, for the purpose of comparing types.

Our solution is rather simple: we split the runtime environment into abstract and concrete parts. Careful readers may have already observed that we use two disjoint alphabets of variable names, one for variables bound in terms $Y$ and one for variables bound in types $Z$. We use $X$ when refering to either kind.
Where the type-specific environments $H$, as discussed above, map only 
term-defined variables to concrete type values created at runtime 
(indexed by $Y$), we use a 
shared environment $J$ (indexed by $Z$), that maps type-defined variables to 
\emph{hypothetical} objects: $\clos{H,T}$ pairs that may or may not correspond 
to an actual object created at runtime.

Implementation-wise, this approach fits quite well with a locally nameless representation of binders \cite{DBLP:journals/jar/Chargueraud12} that already distinguishes between free and bound identifiers.
The runtime subtyping rules for abstract type variables in Figure~\ref{fig:fsub2}
correspond more or less directly to their counterparts in the static subtype relation (Figure~\ref{fig:fsub1}), modulo addition of the two runtime 
environments.

\subsection{Relating Static and Dynamic Subtyping}\label{sec:6subst}

How do we adapt the soundness proof from Section~\ref{sec:stlc} to 
work with this form of runtime subtyping?
First, we need to show that the static typing relation 
implies the dynamic one in well-formed consistent environments:

\infrule[]
{\Gamma \ts T_1 <: T_2 \gap \Gamma \tS H\ J}
{J \ts H\ T_1 <: H\ T_2}
The proof is a simple structural induction 
over the subtyping derivation. Static rules involving $Z$ variables
map to coressponding abstract rules. Rules on $Y$ variables map
to concrete dynamic rules.

Second, for the cases where F$_{\sub}$ type assignment relies on 
substitution in types (specifically, in the result of a type application), 
we need to prove a lemma that enables us to replace a hypothetical binding 
with an actual value:

\infrule[]
{J,Z<:\clos{H,T} \ts H_1\ T_1^Z \sub H_2\ T_2^Z}
{J \ts H_1,Y_1=\clos{H,T}\ T_1^{Y_1} \sub H_2,Y_2=\clos{H,T}\ T_2^{Y_2}}
The proof is again by simple induction and case analysis. We actually
prove a slighty more general lemma that incorporates the option of
weakening, i.e.\ not extending $H_1$ or $H_2$ if $Z$ does not
occur in $T_1$ or $T_2$, respectively.

\subsection{Inversion of Value Typing (Canonical Forms)}\label{sec:6inv}

Due to the presence of the subsumption rule, the main proof can no 
longer just rely on case analysis of the typing derivation, but 
we need proper inversion lemmas for lambda abstractions:

\infrule[]
{H \ts \ v: T_1 \rightarrow T_2}
{v = \clos{H_c,\lambda x:S_1.t} \gap \Gamma_c \tS H_c\ \emptyset\\
\Gamma_c,x:S_1 \ts t: S_2 \gap \emptyset \ts H_c\ S_1 \to S_2 <: H\ T_1 \to T_2
}
And for type abstractions:

\infrule[]
{H \ts \ v: \forall Z<:T_1.T_2^Z}
{v = \clos{H_c,\Lambda Y<:S_1.t} \gap \Gamma_c \tS H_c\ \emptyset\\
\Gamma_c,Y<:S_1 \ts t: S_2^Y \\ 
\emptyset \ts H_c\ \forall Z\!<:\!S_1.S_2^Z\; <:\; H\ \forall Z\!<:\!T_1.T_2^Z
}

We further need to invert the resulting subtyping derivations,
so we need additional inversion lemmas to derive $T_1 <: S_1$ and 
$S_2 <: T_2$ from $S_1 \rightarrow S_2 <: T_1 \rightarrow T_2$ 
and similarly for $\forall$ types.

As already discussed in Section~\ref{sec:2trans} with relation to
DOT, the inversion lemmas we need here depend in a crucial way on 
transitivity and narrowing properties of the subtyping relation
(similar to small-step proofs for F$_{<:}$ \cite{DBLP:conf/tphol/AydemirBFFPSVWWZ05}). 
However, the situation is very different, now that we have
a distinction between runtime values and only static types: as we 
can see from the statements above, we only ever require inversion 
of a subtyping rule in a \emph{fully concrete} dynamic context, 
without any abstract component ($J=\emptyset$)! This means that we 
always have concrete objects at hand, and that we can immediately 
rely on any properties enforced during their construction,
such as `good bounds' in DOT.

\subsection{Transitivity Pushback and Cut Elimination}\label{sec:6trans}

For the static subtyping relation of $F_{<:}$, transitivity can
be proved as a lemma, together with narrowing, in a mutual induction
on the size of the middle type in a chain $T_1 <: T_2 <: T_3$
(see e.g.\ the POPLmark challenge documentation 
\cite{DBLP:conf/tphol/AydemirBFFPSVWWZ05}).

Unfortunately, for the dynamic subtyping version, the same proof
strategy fails, because dynamic subtyping may involve a type variable 
as the middle type: $T_1 <: Y <: T_3$. This setting is very similar
to DOT, but arises surprisingly already when just looking at the
dynamic semantics of $F_{<:}$. Since proving transitivity becomes
much harder, we adopt a strategy from previous DOT developments
\cite{DBLP:conf/oopsla/AminRO14}: admit transitivity as an axiom,
but prove a `pushback' lemma that allows to push uses of the axiom
further up into a subtyping derivation, so that the top level
becomes invertible. We denote this as \emph{precise} subtyping 
$T_1 <!\ T_2$. Such a strategy is reminiscent of cut elimination 
in natural deduction, and in fact, the possibility of cut
elimination strategies is already mentioned in Cardelli's
original $F_{<:}$ paper \cite{DBLP:journals/iandc/CardelliMMS94}.

\paragraph{Cutting Mutual Dependencies}
Inversion of subtyping is only required in a concrete runtime context,
without abstract component ($J = \emptyset$). Therefore, transitivity 
pushback is only required then. Transitivity pushback requires narrowing, 
but only for abstract bindings (those in $J$, never in $H$). 
Narrowing requires these bindings to be potentially imprecise, so 
that the transitivity axiom can be used to inject a step to a
smaller type without recursing into the derivation.
In summary, we need both (actual, non-axiom) transitivity and 
narrowing, but not at the same time. This is a major advantage over
the purely static setting described in Section~\ref{sec:2trans},
where these properties were much more entangled, with no obvious
way to break cycles.

\subsection{Finishing the Soundness Proof}
The interesting case is the one for type application. 
We use the inversion lemma for type abstractions (Section~\ref{sec:6inv}) 
and then invert the resulting subtyping relation on $\forall$ types to relate
the actual type at the call site with the expected type at the
definition site of the type abstraction. In order to do this, 
we invoke pushback once and obtain an invertible subtyping on $\forall$ types. 
But inversion then gives us evidence for the return type in a mixed, 
concrete/abstract environment, with $J$ containing the binding for the 
quantified type variable. Thus we cannot apply transititivy pushback again directly. 
We first need to apply the substitution lemma (Section~\ref{sec:6subst})
to replace the abstract variable reference with a concrete 
reference to the actual type in a runtime environment $H$.
After that, the abstract component is gone ($J = \emptyset$)
and we can use pushback again to further invert the inner 
derivations. 


With that, we have everything in place to complete our soundness
proof for $F_{<:}$:
\infrule[]
{\Gamma \ts e: T \gap \Gamma \tS H \gap \text{eval } n\ H\ e = \text{Done } r}
{r = \text{Val } v \gap v: T \gap}
\vspace{-1ex}

For all $n$, if the interpreter returns a result that is not a timeout, 
the result is also not stuck, and it is well typed.

\subsection{An Alternative Dynamic Typing Relation}\label{sec:6inv-alt}

In the system presented here, we have chosen to include a subsumption rule
in the dynamic typing relation (see Figure~\ref{fig:fsub2}), which required 
us to prove the typing inversion lemmas from Section~\ref{sec:6inv} and
to further handle inversion of subtyping via transitivity pushback 
(Section~\ref{sec:6trans}). 

An alternative design, which we mention for completeness, is to turn this 
around: design the dynamic type assignment relation in such a way that it 
is directly invertible, and prove the subsumption
property (upwards-closure with respect to subtyping) as a lemma. 
To achieve this, we can define type assignment for lambda abstractions
as follows, pulling in the relevant bits of the dynamic subtyping 
relation:

\infrule[]
{\Gamma_c \tS H_c\ \emptyset \gap 
\Gamma_c,x:S_1 \ts t: S_2 \\ 
\emptyset \ts H\ T_1 <: H_c\ S_1 \gap 
\emptyset \ts H_c\ S_2 <: H\ T_2
}
{H \ts \ \clos{H_c,\lambda x:S_1.t}: T_1 \rightarrow T_2}

\noindent
The rule for type abstractions is analogous.

But since we no longer have a built-in subsumption rule, we also
need to pull in the remaining subtyping cases into the 
type assignment relation, such that we can assign $\Top$ and 
concrete type variables $Y$:

\infax[]{H \ts v: \Top}

\infrule[]{H \ts v: T \gap H_2 \ni Y=\clos{H,T}}
{H_2 \ts v: Y}

\noindent
Note, however, that we crucially never assign an abstract type variable $Z$
as type to a value, because value typing is only defined in fully concrete 
environments ($J = \emptyset$, see Figure~\ref{fig:fsub2}).

With these additional type assignment cases the proof of the subsumption 
lemma is straightforward, and dynamic type assignment remains directly
invertible.

In essence, this model performs cut elimination or transitivity pushback 
directly on the type assignment, whereas in Section~\ref{sec:6trans}, we 
performed cut elimination on the subtyping relation. 
Both models lead to valid
soundness proofs for F$_{<:}$ and DOT, but working directly with the 
subtyping relation appears to lead to somewhat stronger auxiliary results.
For this reason, we continue with definitions in the spirit of
Figure~\ref{fig:fsub2} for the remainder of this paper.

\section{From F to D to DOT}\label{sec:dsub}

While we have set out to `only' prove soundness for $F_{<:}$, we have
already had to do much of the work for proving DOT. We will now
extend the calculus to incorporate more DOT features. As we will see,
the changes required to the runtime typing system are comparatively
minor, and in many cases we just remove restrictions of the static
typing discipline.

First of all let us note that $\clos{H,T}$ pairs are already treated 
de-facto as first-class values at runtime. So why not let them loose
and make their status explicit?

\begin{figure}[t]
{\footnotesize$$\ba{llll}
\multicolumn{4}{l}{\text{FSub}\ (\text{F}_{<:})}\\
~&  T &::=& \Top \mid X \mid T \rightarrow T \mid \forall Z<:T.T^Z\\
&  t &::=& x \mid \lambda x:T.t \mid \Lambda Y<:T.t \mid t\ t \mid t\ [T]\\
\\
\multicolumn{4}{l}{\text{DSub}\ (\text{D}_{<:})}\\ 
&  T &::=& \Top \mid x.\text{Type} \mid \{\text{Type}=T\} \mid \{\text{Type}<:T\} \mid (z:T) \rightarrow T^z\\
&  t &::=& x \mid \{\text{Type}=T\} \mid \lambda x:T.t \mid t\ t\\
\\
\multicolumn{4}{l}{\text{DSubBot}}\\
&  T &::=& \Bot \mid  \Top \mid x.\text{Type} \mid \{\text{Type} =T..T\} \mid  (z:T) \rightarrow T^z \\
&  t &::=& x \mid \{\text{Type}=T\} \mid \lambda x:T.t \mid t\ t\\
\\
\multicolumn{4}{l}{\text{DSubBotAndOr}}\\
&  T &::=& \Bot \mid  \Top \mid T \tand T \mid T \tor T \mid x.\text{Type} \\
&    &   & \{\text{Type}=T..T\} \mid (z:T) \rightarrow T^z\\
&  t &::=& x  \mid \{\text{Type}=T\}\mid  \lambda x:T.t \mid t\ t\\
\\
\multicolumn{4}{l}{\text{DSubBotAndOrRec}}\\
&  T &::=& \Bot \mid  \Top \mid T \tand T \mid T \tor T \mid x.\text{Type} \\
&    &   & \{\text{Type}=T..T\} \mid \{l: T\} \mid (z:T) \rightarrow T^z \\
&  t &::=& x \mid \{\text{Type}=T\} \mid \{ \seq{d} \} \mid t.l \mid  \lambda x:T.t \mid t\ t\\
&  d &::=& l=t\\
\\
\multicolumn{4}{l}{\text{DSubBotAndOrRecFix}}\\
&  T &::=& \Bot \mid  \Top \mid T \tand T \mid T \tor T  \mid x.\text{Type} \\
&    &   & \{\text{Type}=T..T\} \mid \{l: T\} \mid (z:T) \rightarrow T^z \mid \{ z \Rightarrow T\} \\
&  t &::=& x \mid \{\text{Type}=T\} \mid \{ \seq{d} \} \mid t.l\mid  \lambda x:T.t \mid t\ t \mid \mu x. t \\
&  d &::=& l=t\\  
\\
\multicolumn{4}{l}{\text{DSubBotAndOrRecFixMut}}\\
&  T &::=& \Bot \mid  \Top \mid T \tand T \mid T \tor T  \mid x.\text{Type} \\
&    &   & \{\text{Type}=T..T\} \mid \{l: T\} \mid (z:T) \rightarrow T^z \mid \{ z \Rightarrow T\} \mid \text{Ref}\ T\\
&  t &::=& x \mid \{\text{Type}=T\} \mid \{ \seq{d} \} \mid t.l\mid  \lambda x:T.t \mid t\ t \mid \mu x. t \mid \text{ref}\ t \mid t \text{!} \mid t \text{:=}\ t\\
&  d &::=& l=t\\  
\\
\multicolumn{4}{l}{\text{DOT}}\\
&  T &::=& \Bot \mid  \Top \mid T \tand T \mid T \tor T \mid x.L \\
&    &   & \{L=T..T\} \mid \{l: T\} \mid m(z:T): T^z \mid \{ z \Rightarrow T\}\\
&  t &::=& x \mid  t.l \mid t.m(t) \mid \{ x \Rightarrow \seq{d} \} \\
&  d &::=& l=t \mid L = T \mid m(x) = t \\
\ea$$}
\caption{\label{fig:steps}Reconstructing DOT bottom-up from F$_{\sub}$}
\end{figure}

To give an intuition what this means, let us take a step back and
consider two ways to define a standard @List@ data type in Scala:
\begin{lstlisting}
class List[E]            // parametric, functional style
class List { type E }    // modular style, with type member
\end{lstlisting}
The first one is the standard parametric version. 
The second one defines the element type @E@ as a type member, 
which can be referenced using a path-dependent type. To see the 
difference in use, here are the two respective signatures of a standard 
@map@ function:
\begin{lstlisting}
def map[E,T](xs: List[E])(f: E => T): List[T] = ...
def map[T](xs:List)(f:xs.E=>T):List & { type E=T } = ...
\end{lstlisting}
Again, the first one is the standard parametric version. The second
one uses the path-dependent type @xs.Elem@ to denote the element
type of the particular list @xs@ passed as argument, and uses
a refined type type @List & { type E=T }@ to define the result
of @map@.

We can already infer from this example above that there must be a relation 
between parametric polymorphism and path-dependent types, an idea that is not 
surprising when thinking of Barendregt's $\lambda$ cube \cite{DBLP:journals/jfp/Barendregt91}. 
But in contrast
to these well-studied systems, DOT adds subtyping to the mix, with 
advanced features like intersection types and recursive types, 
which are not present in the $\lambda$ cube.

But as it turns out, subtyping is an elegant way to model reduction
on the type level. Consider the usual definition of types and terms
in System F \cite{girard1972interpretation,DBLP:conf/programm/Reynolds74}:
$${\ba{llll}
&  T &::=& X \mid T \rightarrow T \mid \forall X.T^X\\
&  t &::=& x \mid \lambda x:T.t \mid \Lambda X.t^X \mid t\ t \mid t\ [T]\\
\ea}$$
We can generalize this system to one with path-dependent types
as well as abstract and concrete type values, which we call System D:
$${\ba{llll}
&  T &::=& x.\text{Type} \mid \{\text{Type}=T\} \mid \{\text{Type}\} \mid (z:T) \rightarrow T^z\\
&  t &::=& x \mid \{\text{Type}=T\} \mid \lambda x:T.t \mid t\ t\\
\ea}$$

System D arises from System F by unifying type and term variables.
References to type variables $X$ are replaced by path-dependent
types $x.\text{Type}$. Type abstractions become term lambdas
with a type-value argument: 
$$
\Lambda X.t^X \gap\leadsto\gap \lambda x:\{\text{Type}\}.t^x
$$
Universal types become dependent function types:
$$
\forall X.T^X \gap\leadsto\gap (x:\{\text{Type}\}) \to T^{x}
$$
And type application becomes dependent function application
with a concrete type value: 
$$
t\ [T] \gap\leadsto\gap t\ \{\text{Type}=T\}
$$
But how do we actually type check such an application?
Let us assume that $f$ is the polymorphic identity function,
and we apply it to type $T$. Then we would like the following
to be an admissible type assignment:
\infrule
{f: (x:\{\text{Type}\}) \to (z: x.\text{Type}) \to x.\text{Type}}
{f\ \{\text{Type}=T\} : (z:T) \to T}
In most dependently typed systems there is a notion of reduction
or normalization on the type level. Based on our definitional interpreter
construction, we observe that we can just as well use subtyping.
For this application to type check using standard
dependent function types we need the following specific
subtyping rules:
\begin{multicols}{2}
\infrule[]
{x: \{\text{Type}=T\}}
{T <: x.\text{Type} \gap x.\text{Type} <: T}

\infax[]{\vspace{1ex}\{\text{Type}=T\} <: \{\text{Type}\}}
\end{multicols}
It is easy to show that System D encodes System F, but not
vice versa. For example, the following function does not
have a System F equivalent: $\lambda x:\{\text{Type}\}.x$

In the following, we are going to reconstruct DOT bottom-up
from System F, based on a series of intermediate calculi, 
shown in Figure~\ref{fig:steps}, of increasing expressiveness. 
The starting point is System F$_{\sub}$, with the full rules 
given in Figure~\ref{fig:fsub1}.
We generalize to System D$_{\sub}$, similar to the exposition above. 
The main difference is that abstract types can be upper bounded. In 
the next step, we enable lower bounding of types, and we can remove 
the distinction between concrete and abstract types. We continue by 
adding intersection and union types, and then records. We add recursive
types, and then mutable references. Finally, we fuse the
formerly distinct notions of functions, records, type values,
and recursive types into a single notion of object with
multiple members. This leaves us with a calculus corresponding
to previous presentations of DOT: a unification of objects,
functions, and modules, corresponding to the rules shown in
Figure~\ref{fig:dot1}.

\subsection{Extension 1: First-Class Type Objects (D$_{<:}$)}\label{sec:fcto}

The first step is to expose first-class type values on the static term level:
$$\ba{lll}
  T &::=& \Top \mid x.\text{Type} \mid \{\text{Type}=T\} \mid \{\text{Type}<:T\} \mid (z:T) \rightarrow T^z\\
  t &::=& x \mid \{\text{Type}=T\} \mid \lambda x:T.t \mid t\ t\\
\ea$$

We present so modified typing rules in Figure~\ref{fig:fsub3}. 
Type objects $\clos{H,T}$ are now first class values, just like
closures. We no longer need two kinds of bindings in $H$ environments,
but we keep the existing structure for $J$ environments, since only 
types can be abstract. We introduce a `type of types', 
$\{\text{Type}=T\}$, and a corresponding introduction term. 
Note that there is no directly corresponding elimination form on
the term level.
References to type variables now take the form $x.\text{Type}$,
where $x$ is a regular term variable--in essence, DOT's path
dependent types but with a unique, global, label @Type@. 
Since type objects are now first class values,
we can drop type abstraction and type application forms and
just use regular lambdas and application, which we extended
to dependent functions and (path) dependent application.

The definitional interpreter only needs an additional case for 
the new type value introduction form:
\begin{lstlisting}[keywords={}]
    | ttyp T => (vty env T)
\end{lstlisting}
Everything else is readily handled by the existing
lambda and application cases.

In the typing rules in Figure~\ref{fig:fsub3}, we have a new
case for type values, invariant in the embedded type. The
rules for type selections (previously, type variables) are
updated to require bindings in the environment to map to 
$\{\text{Type}=T\}$ types. 

\begin{figure}[h!]
\begin{framed}
\judgement{Syntax}{}

\medskip


$\ba{lll}
  T &::=& x.\text{Type} \ |\  \Top \ |\ (z:T) \rightarrow T^z \ |\ \{\text{Type}=T\}\\
  t &::=& x \ |\  \lambda x:T.t \ |\ t\ t \ |\ \{\text{Type}=T\}\\
  \Gamma &::=& \emptyset \ |\ \Gamma,x:T\\[1ex]
  v &::=& \clos{H,\lambda x:T.t} \ |\ \clos{H,T}\\
  H &::=& \emptyset \ |\ H,x:v\\
  J &::=& \emptyset \ |\ H,z:\clos{H,T}
\ea$

\medskip

\judgement{Subtyping}{\fbox{$\Gamma \ts S \sub U$}}

  \infrule[]{\Gamma \ts T_1 \sub T_2 \gap \Gamma \ts T_2 \sub T_1}
  {\Gamma \ts \{\text{Type}=T_1\}\ \sub\ \{\text{Type}=T_2\}}

  \infrule[]{\Gamma(x) = U \gap \Gamma \ts U \sub \{\text{Type}=T\}}
  {\Gamma \ts x.T \sub T}

  $\ldots$

\judgement{Type assignment}{\fbox{$\Gamma \ts t : T$}}

  \infax[]{\Gamma \ts \{\text{Type}=T\}: \{\text{Type}=T\}}

  \infrule[]{\Gamma,x:T_1 \ts t_2: T_2^x}
  {\Gamma \ts \lambda x:T_1.t_2: (z:T_1)\rightarrow T_2^z}

  \infrule[]{\Gamma \ts t_1: (z:T_1)\rightarrow T_2^z  \spcomma t_2: T_1}
  {\Gamma \ts t_1 t_2: T_2}

  $\ldots$

\judgement{Runtime Subtyping}{\fbox{$J \ts H_1\ T_1 \sub H_2\ T_2$}}

  \infrule[]{J \ts H_1\ T_1 \sub H_2\ T_2  \gap J\ts H_1\ T_2 \sub H_2\ T_1}
  {J \ts H_1\ \{\text{Type}=T_1\}\ \sub\ H_2\ \{\text{Type}=T_2\}}

  \infrule[]{J(z) = \clos{H,U} \gap J \ts H\ U \sub H_2\ \{\text{Type}=T\}}
  {J \ts H_1\ z.\text{Type} \sub H_2\ T}

  \infrule[]{H_1(x)\!=\!v \;\; \ H\!\ts\!v\!: U \gap J \ts H\ U \sub H_2\ \{\text{Type}=T\}}
  {J \ts H_1\ x.\text{Type} \sub H_2\ T}

  $\ldots$

\judgement{Value type assignment}{\fbox{$H \ts v : T$}}

  \infrule[]{\Gamma \tS H\ \emptyset \gap \Gamma \ts \{\text{Type}=T\}: \{\text{Type}=T\}}
  {H \ts \clos{H,T}: \{\text{Type}=T\}}

  \infrule[]{\Gamma \tS H \emptyset \gap \Gamma,x:T_1 \ts t: T_2^x}
  {H \ts \clos{H,\lambda x:T_1.t}: (z:T_1)\rightarrow T_2^z}
          
\caption{\label{fig:fsub3}D$_{\sub}$: generalizing F$_{\sub}$ with first-class types}
\end{framed}
\end{figure}

\subsection{Another Level of Transitivity Pushback}

The generalization in this section is an essential step towards DOT.
Most of the changes to the soundness proof are rather minor.
However, one piece requires further attention: the previous
transitivity pushback proof relied crucially on being able
to relate types across type variables:
$$H_1\ T_1 <!\ H\ Y <!\ H_3\ T_3$$
Inversion of this derivation would yield another chain
$$H \ni \clos{H_2,T_2} \gap H_1\ T_1 <:\ H_2\ T_2 <:\ H_3\ T_3,$$
which, using an appropriate induction strategy, can be
further collapsed into $H_1 <!\ H_3$.

But now the situation is more complicated: inversion of
$$H_1\ T_1 <!\ H\ x.\text{Type } <!\ H_3\ T_3$$
yields
$$ H(x) = v \gap H_2 \ts v: T_2$$
$$ H_2\ T_2 <:\ H_1\ \{\text{Type}=T_1\} \gap
   H_2\ T_2 <:\ H_3\ \{\text{Type}=T_3\}, $$
but there is no immediate way to relate $T_1$ and $T_3$!
We would first have to invert the subtyping relations
with $T_2$, but this is not possible because these relations 
are imprecise and may use transitivity. Recall that they have
to be, because they may need to be 
narrowed---but wait! Narrowing is only required for
abstract types, and we only need inversion and transitivity
pushback for fully concrete contexts. So, while the imprecise
subtyping judgement is required for bounds initially,
in the presence of abstract types, we can replace it with 
the precise version once we move to a fully concrete
context.

This idea leads to a solution involving another pushback
step. We define an auxiliary relation $T_1 <<: T_2$, which 
is just like $T_1 <: T_2$, but with precise lookups. For
this relation, pushback and inversion work as before,
but narrowing is not supported. To make sure we
remain in fully concrete contexts only, where we do not need
narrowing, we delegate to $<:$ in the body of the dependent 
function rule:

\infrule[]{J \ts H_2\ T_1 \sub H_1\ S_1\\
J,z:\clos{H_2,T_1} \ts H_1\ S_2^z \sub H_2\ T_2^z}
{J \ts H_1\ \lambda x:S_1.S_2^x\ <<!\ \ H_2\ \lambda x:T_1.T_2^x}

In this new relation, we can again remove top-level uses of the 
transitivity axiom. A derivation $T_1 <: T_2$ can be converted into 
$T_1 <<: T_2$ and then further into $T_1 <<!\ T_2$.
With that, we can again perform all the necessary inversions 
required for the soundness proof.

\subsection{Extension 2: Subtyping Lattice and Records (DSubBotAndOrRec)}

Abstract types in F$_{\sub}$ can only be bounded from above. 
But internally, the runtime typing already needs to support type
symmetric rules, for type selections on either side. We can easily
expose that facility on the static level as well. To do so, we
add a bottom type $\bot$ and extend our type values to include
both lower and upper bounds, as in DOT: $\{\text{Type}=S..U\}$.
$$\ba{lll}
  T &::=& ... \mid \Bot \mid T \tand T \mid T \tor T \mid \{\text{Type}=T..T\}\\
\ea$$

What is key is that in any partially 
abstract context, such as when comparing two dependent function
types, lower and upper bounds for an abstract type need not be 
in any relation. This is key for supporting DOT, because narrowing 
in the presence of intersection types may turn perfectly fine, 
`good bounds' into contradictory ones (Section~\ref{sec:goodbounds}).

However, we do check for good bounds at object creation 
time, so we cannot ever create an object with bad bounds. 
In other words, any code path that was type checked using 
erroneous bounds will never be executed. With these facilities
in place, we can add intersection and union types without much trouble.

We also introduce records:
$$\ba{lll}
  T &::=& ... \mid \{l: T\} \\
  t &::=& ... \mid \{ \seq{d} \} \mid t.l \\
  d &::=& l=t\\
\ea$$

The type of a record will be an intersection type corresponding
to its fields.

The modifications in this Section are all rather straighforward,
so we elide a formal presentation of the modified rules for
space reasons.

\subsection{Extension 3: Recursive Self Types (DSubBotAndOrRecFix)}

A key missing bit to reach the functionality of DOT is support
for recursion and recursive self types.

$$\ba{lll}
  T &::=& ... \mid \{ z \Rightarrow T\}\\
  t &::=& ... \mid \mu x. t\\
\ea$$

Recursive self types enable type members within
a type to refer to one another. Similar to $\forall$ types, 
they introduce a new type variable binding, 
and they require narrowing, transitivity pushback, etc.

Recursive self types are somewhat similar to existentials. However 
they cannot readily be encoded in F$_{\sub}$. In DOT, they 
are also key for encoding refinements, together with intersection
types (see Section~\ref{sec:main1}).
Self types do not have explicit pack and unpack operations, 
but any variable reference can pack/unpack self types
on the fly (rules from Figure~\ref{fig:dot1}):

\bigskip
\begin{minipage}{1.8cm}
  \infrule[]{\Gamma(x) = T}
  {\Gamma \ts x: T}
\end{minipage}
\begin{minipage}{2.2cm}
  \infrule[]{\Gamma \ts x: T^x}
  {\Gamma \ts x: \{z \Rightarrow T^z\}}
\end{minipage}\qquad
\begin{minipage}{3cm}
  \infrule[]{\Gamma_{[x]} \ts x: \{z \Rightarrow T^z\}}
  {\Gamma \ts x: T^x}
\end{minipage}

\bigskip

In full DOT, we assign recursive self types at object construction 
(see Figure~\ref{fig:dot1}). Here, we do not have a notion of objects yet. 
Instead, we provide an explicit fixpoint combinator $\mu x. t$, with
a standard implementation. The static type rule is as follows:

\infrule[]
{\Gamma,x:T^x \ts t: T^x}
{\Gamma \ts \mu x. t: \{ z \Rightarrow T^z\}}

In the premise, we can always apply the pack rule to assign $x$ 
type $\{ z \Rightarrow T^z\}$.

To assign it a type, we first look at the context 
with the object itself bound to a fresh identifier. Then we apply 
the pack rule to that identifier to assign a self type.

Enabling unpacking of self types in subtyping is considerably 
harder.

\subsection{Pushback, Once More}\label{sec:oncemore}

So far, the system was set up carefully to avoid cycles between 
required lemmas. Where cycles did occur, as with transitivity and
narrowing, we broke them using a pushback technique.
A key property of the system is that, in general, we are very 
lenient about things outside of concrete runtime contexts.
The only place where we invert a $\forall$ type or dependent
function type and go from hypothetical to concrete is in showing 
safety of the corresponding type assignment rules. This enables
subtyping inversion and pushback to disregard abstract binding
for the most part.

When seeking to unpack self types within lookups of type
selections in subtype comparisons, these assumptions are no
longer valid. Every lookup of a variable, while considering a path 
dependent type, may potentially need to unfold and invert self 
types. In particular, the pushback lemma itself that converts 
imprecise into precise bounds may unfold a self type. Then
it will be faced with an abstract variable that
first needs to be converted to a concrete value.
More generally, whenever we have a chain 
$$ \{z \Rightarrow T_1\} <: T <: \{z \Rightarrow T_2\},$$ 
we first need to apply transitivity pushback to perform inversion. 
But then, the result of inversion will yield another imprecise
derivation
$$ T_1 <: U <: T_2$$ 
which may be bigger than the original derivation due to 
transitivity pushback. So, we cannot process the result of 
the inversion further during an induction.
This increase in size is a well-known property of cut elimination:
removing uses of a cut rule (like our transitivity axiom) from a proof
term may increase the proof size exponentially.

We solve this issue by yet another level of pushback, 
which this time also needs to work
in partially abstract contexts. The key idea is to
pre-transform all unpacking operations on concrete
values, so that after inversion, only the previous
pushback steps are necessary. We define the
runtime subtyping rules with unpacking as 
follows:

\infrule[]{H_1(x)\!=\!v, \; H_c\!\ts\!v\!:T_c \gap \emptyset \ts H_c\ T_c \sub H_2\ \{z \Rightarrow L:\bot..U\}}
{J \ts H_1\ x.L \sub H_2\ U}
\vspace{-1ex}

The fact that $J=\emptyset$ in the premise is of key importance here,
as it enables us to invoke the previous pushback level to $<<:$ and
finally to $<<!$ even though we are in a partially abstract context
when we traverse a larger derivation.

This restriction to $J=\emptyset$ delivers the explanation for
the use of $\Gamma_{[x]}$ in the premise of the var unpacking
rule in Figure~\ref{fig:dot1}.

\subsection{Mutable References (DSubBotAndOrRecFixMut)}

As a further extension, we add ML-style mutable references. 
$$\ba{lll}
  T &::=& \dots \mid \text{Ref}\ T\\
  t &::=& \dots \mid \text{ref}\ t \mid t \text{!} \mid t \text{:=}\ t \\
  v &::=& \dots \mid \text{loc}\ x
\ea$$

The extension
of the syntax and static typing rules is standard, with a new syntactic
category of store locations. The evaluator is augmented with a runtime
store, and reading or writing to a location accesses the store.
How do we assign a runtime type to a store location? The key difficulty
is that store bindings may be recursive, which has lead Tofte to discover
coinduction as a proof technique for type soundness \cite{Tofte88operationalsemantics}.
We sidestep this issue by assigning types to values (in particular store 
locations) with respect to a \emph{store typing} $S$ instead of the store itself.
Store typings consist of $H\ T$ pairs, which can be related through the usual 
runtime subtyping judgements. The value type assignment judgement now 
takes the form $S\ H \ts v: T$ and since subtyping depends on
value type assignment, it is parameterized by the store typing as well:
$S\ J \ts H_1\ T_1 <: H_2\ T_2$. The type assignment rule for store locations
simply looks up the correct type from the store typing:

\infrule[]
{S(x) = \clos{H,T}}
{S\ H \ts \text{loc}\ x: \text{Ref}\ T}

When new bindings are added to the store, they are assigned the
type and environment from their creation site in the store typing. 
When accessing the store, bindings in the store typing are always 
preserved, i.e.\ store typings are invariant under reads and updates. 

Objects in the store must conform to the store typing at all times. 
With that, an update only has to provide a subtype of the type in the 
store typing, and it will not change the type of that slot. So if an update 
creates a cycle in the store, this does not introduce circularity in the store typing.

A canonical forms lemma for references states that if a value $v$ has 
type $\text{Ref}\ T$, $v$ must be a store location with a type equal 
to $T$ in the store typing.

The main soundness statement is modified to guarantee that if evaluation 
terminates, it produces a well-typed value, and a store that corresponds 
to an extension of the initial store typing. Thus, the store typing is
required to grow linearly, while the values in the store may change 
arbitrarily within the constraints given by the store typing.

We believe that the ease with which we were able to add mutable
references is a further point in favor of definitional interpreters.
Back in 1991, the fact that different proof techniques were thought to
be required to support references in big-step style was a major criticism 
by Wright and Felleisen and a problem their syntactic approach sought
to address \cite{DBLP:journals/iandc/WrightF94}. 


\subsection{The Final DOT Soundness Result}

As the final step, we unify the separate constructs for recursion, 
records, and lambda abstractions, into a single notion of \emph{object}.

$$\ba{lll}
T &::=& \Bot \mid  \Top \mid T \tand T \mid T \tor T \mid x.L \\
  &   & \{L=T..T\} \mid \{l: T\} \mid m(z:T): T^z \mid \{ z \Rightarrow T\}\\
t &::=& x \mid  t.l \mid t.m(t) \mid \{ x \Rightarrow \seq{d} \} \\
d &::=& l=t \mid L = T \mid m(x) = t \\
\ea$$

Objects can have type, value, and method members with distinct labels.
We could also add mutable fields based on the handling of references above
but we disregard this option for simplicity.

Integrating these features does not pose particular difficulties.
Some aspects are even simplified, for example the treatment of
recursion. Instead of having to support an explicit fixpoint
combinator, we can implement recursion by passing the @this@
object on which a method is called as an additional silent
parameter to the method.

With all this, we obtain our final soundness result for DOT:
\infrule[]
{\Gamma \ts e: T \gap \Gamma \tS H \gap \text{eval } n\ H\ e = \text{Done } r}
{r = \text{Val } v \gap v: T \gap}
\vspace{-1ex}

For all $n$, if the interpreter returns a result that is not a timeout, 
the result is also not stuck, and it is well typed. 



We have seen how DOT emerges from F$_{<:}$ through relatively gentle extensions 
which we identify as variations of D$_{<:}$. 
This naturally raises the question what other interesting type systems can arise by 
devising suitable static rules from the dynamic ones given by the
interpreter and environment structure. We believe this is an exciting 
new research angle.

\section{Back to Small-Step Proofs}\label{sec:smallstep}

While our approach of using big-step evaluators has lead to important 
insights and produced the first soundness proof for DOT,
the results can also be transferred to a small-step setting.

The general idea is to preserve the distinction between static type
assignment to \emph{terms} and dynamic type assignment to \emph{values}. 
A key obstacle is that our definitional interpreter uses environments
whereas our target reduction semantics will be based on substitution.
As we have seen in Section~\ref{sec:nosubst}, naive substitution of values
for identifiers has no chance of working because it would break the 
structure of path-dependent types.

As a way out, we first introduce an additional level of indirection
in our interpreter. Instead of storing values directly in environments, we
allocate all objects in a store-like structure, which grows monotonically.
In this setting, only store locations can be first-class values, i.e.\ 
passed around between functions and stored in environments. 
This additional level of indirection does not interfere with evaluation 
and clearly preserves the semantics of the interpreter.

But now the use of environments is no longer necessary: since we are
dealing only with pointers into our auxiliary store data structure, 
we can just as well use substitution instead of explicit environments. 
For runtime subtyping, the two-handed judgement 
$$
  J \ts H_1\ T_1 <: H_2\ T_2 
$$
where $J$ holds bindings from self-type comparisons and $H_1$,$H_2$ hold 
runtime values becomes
$$
 H\ J \ts T_1 <: T_2
$$
where $J$ remains as it is, and free names from $T_1$ and $T_2$ are 
now resolved in the shared store $H$.

We can mechanically transform this substitution-based evaluator
into a substitution-based reduction semantics by following the techniques
of Danvy et al.\ \cite{DBLP:journals/jcss/DanvyJ10,DBLP:journals/tcs/DanvyMMZ12,DBLP:conf/ppdp/AgerBDM03}. 
We transform the interpreter to continuation passing style (CPS) and 
defunctionalize the continuations to arrive at a first small-step semantics in 
form of an abstract machine. After exploiting this functional correspondance
between evaluators and abstract machines, we exploit the syntactic 
correspondance between abstract machines and reduction semantics,
and obtain our desired term rewriting rules.

We can represent the store syntactically as a sequence of @let-store@ 
bindings. The key reduction rule that allocates type values as new 
identifiers in the store can be phrased as follows (example from D$_{<:}$):
\infax[]
{E [ \{\textrm{Type} = T\} ]  \longrightarrow  
\texttt{let-store}\ x = \{\textrm{Type} = T\} \texttt{in}\ E[ x ]}
Where $E$ is a @let-store@-free evaluation context and $x$ a fresh
name. Thus, substitution in 
path-dependent types occurs only with store locations @x@.

Where the big-step evaluator had separate static and dynamic notions
of subtyping, it is beneficial to combine them into one judgment in
small-step, but we need to retain different behavior for path-dependent
types depending on whether the variable is bound to a value in the store, 
to an identifier bound in a term or to an identifier introduced in a
subtype comparison. The combined subtyping judgement thus takes the 
following form:
$$
 H\ G\ J \ts T_1 <: T_2
$$
where $H$ contains the store bindings (i.e.\ runtime values), $G$ term bindings, 
and $J$ bindings from sub-type comparisons. Potentially, $J$ and $G$ can 
be merged into one.

With this, the same approach as in our big step model applies: we can freely 
narrow within $J$ because we do not need to check bounds, and we can just use 
the transitivity axiom. When we substitute a term variable with a store 
location (i.e.\ when we go from $J$ or $G$ to $H$), we know that type 
bounds are well-formed, so top-level uses of the transitivity axiom 
can be eliminated using push-back techniques (Section~\ref{sec:6trans}).

We treat only store locations as first-class values. For assigning types 
to store locations (and the objects they map to), inversion lemmas, and
related issues, all the considerations from 
Sections~\ref{sec:6inv} to \ref{sec:6inv-alt} apply.

\section{Related Work}\label{sec:related}

\paragraph{Scala Foundations} Much work has been done on grounding 
Scala's type system in theory. Early efforts included $\nu$Obj~\cite{nuObj}, 
Featherweight Scala~\cite{FS} and Scalina~\cite{scalina}, all of them 
more complex than what is studied here. None of them lead to
a mechanized soundness result, and due to their inherent complexity, 
not much insight was gained why soundness was so hard to prove. 
DOT \cite{dotfool} was proposed as a simpler and more foundational 
core calculus, focusing on path dependent types but disregarding classes,
mixin linearization and similar questions. The original DOT formulation 
\cite{dotfool} had actual preservation issues because lookup was required 
to be precise. This prevented narrowing, as explained in section
\ref{sec:main3}. The originally proposed small step rewriting semantics 
with a store exposed the difficulty of relating paths at different 
stages of reductions. 

The $\mu$DOT calculus \cite{DBLP:conf/oopsla/AminRO14} 
is the first calculus in the line with a mechanized soundness result,
(in Twelf, based on total big step semantics), but the calculus is much 
simpler than what is studied in this paper. Most importantly, $\mu$DOT 
lacks bottom, intersections and type refinement. 
Amin et al.\ \cite{DBLP:conf/oopsla/AminRO14} describe in great detail 
why adding these features causes 
trouble. Because of its simplicity, $\mu$DOT supports both narrowing 
and transitivity with precise lookup. The soundness proof for $\mu$DOT
was also with respect to big-step semantics. However, the semantics
had no concept of distinct runtime type assignment and would thus 
not be able to encode F$_{<:}$ and much less full DOT.

After the first version of this paper was circulated as a tech 
report, a small-step semantics and soundness proof sketch for a DOT
variant close to the one described here was proposed by 
Odersky. It has recently been mechanized and accepted for 
publication \cite{wadlerfest}. While on the surface 
similar to the DOT version presented here, there are 
some important differences. First, the calculus in \cite{wadlerfest} 
is restricted to Administrative Normal Form (ANF) \cite{DBLP:conf/pldi/FlanaganSDF93}, 
requiring all intermediate subexpressions to be let-bound with explicit names.
Second, the calculus does not support subtyping between 
recursive types, only their introduction and elimination as part 
of type assignment. This skirts the thorniest issues in the proofs
(see Section~\ref{sec:oncemore}) but also limits the expressiveness
of the calculus. For example, an identifier $x$ bound
to a refined type $\{ z \Rightarrow T \tand U^z \}$ can be 
treated as having type $T$, but if it instead has type
$S \to \{ z \Rightarrow T \tand U^z \}$, it can not be assigned
type $S \to T$. Instead, one has to eta-expand the term into
a function, let-bind the result of the internal call, 
and insert the required coercion to $T$. 
Similar considerations apply to types in other non-toplevel 
positions such as bounds of type members, but it is not
clear if an analogue of eta-expansion is always available. 
Third, the small-step proof in \cite{wadlerfest}
is presented without any formal connection to the earlier
definitional interpreter result.
The present revision of this paper lifts all these restrictions, 
by providing a mechanized small-step proof that is not restricted 
to ANF, supports full subtyping between recursive types, and is 
constructed in a systematic way from the definitional 
interpreter result (Section~\ref{sec:smallstep}).

\paragraph{ML Module Systems}
1ML~\cite{DBLP:conf/icfp/Rossberg15} unifies the ML module and core
languages through an elaboration to System F$_\omega$ based on earlier
such work~\cite{DBLP:journals/jfp/RossbergRD14}. Compared to DOT, the 
formalism treats recursive modules in a less general way and it
only models fully abstract vs fully concrete types, not bounded 
abstract types. 
Although an implementation is provided, there is no mechanized proof. 
In good ML tradition, 1ML supports Hindler-Milner style type inference, 
with only small restrictions. 
Path dependent types in ML modules go back at least
to SML~\cite{Macqueen86usingdependent}, with foundational work on
transparent bindings by Harper and Lillibridge~\cite{homl} and 
Leroy~\cite{leroy:manifest}. MixML~\cite{mixml} drops the
stratification requirement and enables modules as first class
values.

\paragraph{Other Related Languages}
Other languages and calculi that include features related to
DOT's path dependent types include the family polymorphism
of Ernst~\cite{DBLP:conf/ecoop/Ernst01}, 
Virtual Classes~\cite{vc,conf/ecoop/Ernst03,DBLP:conf/oopsla/NystromCM04,conf/oopsla/GasiunasMO07},
and ownership type systems like Tribe~\cite{tribe,tribalo}.
Nomality by ascription is also achieved in Grace~\cite{DBLP:conf/ecoop/JonesHN15}.

\paragraph{Semantics and Proof Techniques}

There is a vast body of work on soundness and proof techniques.
The most relevant here is Wright and Felleisen's syntactic
approach \cite{DBLP:journals/iandc/WrightF94}, Kahn's Natural Semantics
\cite{DBLP:conf/stacs/Kahn87}, and Reynold's Definitional Interpreters
\cite{DBLP:journals/lisp/Reynolds98a}. We build our proof technique
on Siek's Three Easy Lemmas \cite{siek3lemmas}. Other work that
discusses potential drawbacks of term rewriting techniques
includes Midtgaard's survey of interpreter implementations 
\cite{DBLP:conf/ppdp/MidtgaardRL13}, 
Leroy and Grall's coinductive natural
semantics \cite{DBLP:conf/esop/NeronTVW15} and Danielsson's
semantics based on definitional interpreters with coinductively
defined partiality monads. Coinduction also was a key enabler 
for Tofte's big-step soundness proof of core ML 
\cite{Tofte88operationalsemantics}. In our setting, we get 
away with purely inductive proofs, thanks to numeric step
indexes or depth bounds, even for mutable
references. We believe that ours is the first purely inductive
big-step soundness proof in the presence of mutable state. 
Step counters also play a key role in proofs based on
logical relations \cite{DBLP:conf/esop/Ahmed06}.
Our runtime environment 
construction bears some resemblance to Visser's name graphs
\cite{DBLP:conf/esop/NeronTVW15} and also to Flatt's bindings as 
sets of scopes \cite{DBLP:conf/popl/Flatt16}.
Big-step evaluators can be mechanically transformed into equivalent
small-step semantics following the techniques of
Danvy et al.\ \cite{DBLP:journals/jcss/DanvyJ10,DBLP:journals/tcs/DanvyMMZ12,DBLP:conf/ppdp/AgerBDM03}.

\section{Conclusions}\label{sec:conclusion}

We have presented a soundness result for a variant of DOT
that includes type refinement and a subtyping lattice with
full intersection types,
demonstrating how the difficulties that prevented such a
result previously can be overcome with a semantic model
that exposes a distinction between static terms and runtime 
values.

Along the way, we have presented the first type soundness proof for 
System F$_{<:}$ that is based on a high-level functional evaluator 
instead of a term rewriting system, establishing that the
`definitional interpreters approach to type soundness' scales 
to sophisticated polymorphic type systems of broad interest.

By casting DOT as an extension of the runtime behavior of 
F$_{<:}$ to the static term level, we have exposed new insights
into the design of DOT-like calculi in particular, with
intermediate systems such as F$_{<:}$, and a new avenue for 
the exploration and design of type systems in general.

\section*{Acknowledgements}

The initial design of DOT is due to Martin Odersky.
Geoffrey Washburn, Adriaan Moors, Donna Malayeri, Samuel Grütter 
and Sandro Stucki have contributed to previous or alternative developments.
For insightful discussions we thank Amal
Ahmed, Jonathan Aldrich, Derek Dreyer, Sebastian Erdweg, Erik Ernst, 
Matthias Felleisen, Ronald Garcia, Paolo Giarrusso, Scott Kilpatrick, 
Grzegorz Kossakowski, Alexander Kuklev, Viktor Kuncak, Alex Potanin, 
Jon Pretty, Didier R\'emy, Lukas Rytz, Miles Sabin, Ilya Sergey, 
Jeremy Siek, Josh Suereth, Ross Tate, Eelco Visser, and Philip Wadler.

{
\bibliographystyle{abbrv}
\bibliography{dot}
}

\end{document}